\documentclass[a4paper,twoside]{article}

\usepackage[left=2cm, right=2cm, top=2cm, bottom=2cm]{geometry}
\usepackage{float}
\usepackage{epsfig}
\usepackage{subcaption}
\usepackage{calc}
\usepackage{amssymb}
\usepackage{amstext}
\usepackage{amsmath}
\usepackage{amsthm}
\usepackage{units}
\usepackage{multicol}
\usepackage{pslatex}
\usepackage{apalike}
\usepackage{algorithm2e}
\usepackage[bottom]{footmisc}
\usepackage{dblfloatfix}
\usepackage{authblk}

\usepackage{tikz}
\usetikzlibrary{datavisualization,calc,intersections,through,backgrounds,math}
\usepackage{pgfplots}
\pgfplotsset{width=7cm,compat=1.16}
\usepgfplotslibrary{fillbetween}
\pgfplotsset{compat=1.16}
\usetikzlibrary{positioning}
\usetikzlibrary{fadings}
\usetikzlibrary{patterns}
\usetikzlibrary{pgfplots.patchplots}
\usetikzlibrary{decorations.text}
\usetikzlibrary{perspective}
\usetikzlibrary{arrows.meta}
\usetikzlibrary{shadows}

\definecolor{myred}{rgb}{.906,.161,.541}
\definecolor{myblue}{rgb}{.459,.439,.702}
\definecolor{mygreen}{rgb}{.106,.620,.467}
\definecolor{myorange}{rgb}{.851,.373,.008}

\newcommand\ie{i.\,e., }
\newcommand\eg{e.\,g., }
\newcommand\etal{et al.\ }
\def\twod/{2\protect\nobreakdash-d}
\def\threed/{3\protect\nobreakdash-d}
\newcommand\Figure{Figure~}
\newcommand\Section{Section~}

\newcommand*\diffm{\mathop{}\!\mathrm{d}}

\newcommand*\restrr[2]{{
  \left.\kern-\nulldelimiterspace
  #1
  \mathclose{}
  \vphantom{\big|}
  \right|_{#2}
  }}

\newcommand*\coeffoff[2]{{
  \left[#1\right]
  \mathclose{}
  \mathopen{}
  #2
  }}
\newcommand*\coeffoffk[3]{{
  \left[#1\right]_{#2}
  \mathclose{}
  \mathopen{}
  #3
  }}
  \global\long\def\coeffofk#1#2#3{\coeffoffk{\knotPosition^{#1}}{#2}{#3}}

\begin{document}

\title{Particle-Wise Higher-Order SPH Field Approximation for DVR}
\date{}
\renewcommand\Authfont{\small}
\author[ \hspace{-1ex}]{Jonathan Fischer}
\author[ \hspace{-1ex}]{Martin Schulze}
%\affil[*]{unaffiliated}
\author[1]{Paul Rosenthal}
\author[2]{Lars Linsen}
\affil[1]{Institute for Visual and Analytic Computing, University of Rostock, Albert-Einstein-Str. 22, 18059 Rostock, Germany}
\affil[2]{Institute of Computer Science, University of Münster, Einsteinstr. 62, 48149 Münster, Germany}

\maketitle

\abstract{
\noindent When employing Direct Volume Rendering (DVR) for visualizing volumetric scalar fields, classification is generally performed on a piecewise constant or piecewise linear approximation of the field on a viewing ray.
Smoothed Particle Hydrodynamics (SPH) data sets define volumetric scalar fields as the sum of individual particle contributions, at highly varying spatial resolution.
We present an approach for approximating SPH scalar fields along viewing rays with piecewise polynomial functions of higher order. This is done by approximating each particle contribution individually and then efficiently summing the results, thus generating a higher-order representation of the field with a resolution adapting to the data resolution in the volume.
\newline

}
\vspace{1cm}
\begin{multicols}{2}
\global\long\def\on#1{\operatorname{#1}}%

\global\long\def\v#1{\boldsymbol{#1}}%

\global\long\def\diff{\diffm}%

\global\long\def\fc#1#2{\mathop{#1}\mathopen{}\left(#2\right)}%

\global\long\def\restr#1#2{\restrr{#1}{#2}}%

\global\long\def\indentEqual{\hphantom{\mathord{}=\mathord{}}}%

\global\long\def\indentPlus{\hphantom{\mathord{}+\mathord{}}}%

\global\long\def\indentDot{\hphantom{\mathord{}\cdot\mathord{}}}%

\global\long\def\firstPosNatNumbers#1{\left\{  1,\dots,#1\right\}  }%

\global\long\def\intmax{\text{INT{\_}MAX}}

\global\long\def\representative#1{#1_\text{repr}}%

\global\long\def\indexedField#1#2{#1^{\left[#2\right]}}%

\global\long\def\proj#1#2{\fc{\on{proj}_{#1}}{#2}}%

\global\long\def\lastKnotIndex{\left\lceil \nicefrac{\vphantom{K}\smash{\numpapprox}}{2}\right\rceil }%

\global\long\def\indicator#1{\boldsymbol{1}_{#1}}%

\global\long\def\coeffof#1#2{\coeffoff{#1}{#2}}%

\global\long\def\round#1{\left\lceil #1\right\rfloor }%

\global\long\def\mortonLess{\le_{{\scriptscriptstyle Z}}}%

\global\long\def\halfIntegerArrayA{\hat{\boldsymbol{a}}}%

\global\long\def\halfIntegerA{\hat{a}}%

\global\long\def\pwpolA{A}%

\global\long\def\pwpolAPiece#1{A_{#1}}%

\global\long\def\pwpolACoeff{a}%

\global\long\def\pwpolAmax{\pwpolACoeff_{\max}}

\global\long\def\leafA{\mathfrak{a}}%

\global\long\def\fieldA{A}%

\global\long\def\setA{\mathcal{A}}%

\global\long\def\innerLeavesSetA{\mathfrak{A}}%

\global\long\def\firstNatNumbersSet{\mathcal{A}}%

\global\long\def\squareIntegrableA{A}%

\global\long\def\orthogbasiselement#1#2{A_{#1#2}}%

\global\long\def\nonorthogbasiselement#1#2{\tilde{A}_{#1#2}}%

\global\long\def\meca{A_{\mergeIndexSet}}%

\global\long\def\mecamax{A_{\max,\mergeIndexSet}}%

\global\long\def\orthogbasis{\mathcal{A}}%

\global\long\def\nonorthogbasis{\tilde{\mathcal{A}}}%

\global\long\def\halfIntegerB{\hat{b}}%

\global\long\def\halfIntegerArrayB{\hat{\boldsymbol{b}}}%

\global\long\def\leafB{\mathfrak{b}}%

\global\long\def\squareIntegrableB{B}%

\global\long\def\setB{\mathcal{B}}%

\global\long\def\firstNatNumbersSubset{\mathcal{B}}%

\global\long\def\firstNatNumbersSubsetElement{b}%

\global\long\def\rayBasePoint{\v b}

\global\long\def\fieldB{B}%

\global\long\def\mecb{B_{\mergeIndexSet}}%

\global\long\def\ksct#1{B_{#1}}%

\global\long\def\searchAreaCenter{\v c}%

\global\long\def\halfIntegerArrayC{\hat{\boldsymbol{c}}}%

\global\long\def\halfIntegerC{\hat{c}}%

\global\long\def\meof{C_{\mergeIndexSet}}%

%\global\long\def\degw{D}%

\global\long\def\degapprox{D}%

\global\long\def\kpidomain{\mathcal{D}}%

%\global\long\def\mergeError#1{E_{#1}}%

%\global\long\def\twoParticlesMergeError#1#2{\fc E{#1,#2}}%

\global\long\def\particlesBestMergeError#1{\fc{\check{E}}{#1}}%

\global\long\def\particlesOuterMergeError#1{\fc{\hat{E}}{#1}}%

\global\long\def\innerLeafBME#1{\fc{\check{E}}{#1}}%

\global\long\def\innerLeavesSetBME#1{\fc{\hat{E}}{#1}}%

\global\long\def\approxError{E_{\adst}}%

\global\long\def\bestApproxError#1{E_{#1}}%

\global\long\def\bestBestApproxError#1{\bestApproxError{#1}^{*}}%
\global\long\def\bestBestApproxErrorSquared#1{\bestApproxError{#1}^{*\;2}}%

\global\long\def\overallApproxErrorSpecific#1#2{E_{#1,#2}}%
\global\long\def\overallApproxError{\overallApproxErrorSpecific{\numpapprox}{\degapprox}}%
\global\long\def\overallApproxErrorSquared{E^{*\;2}}%

\global\long\def\rNatIndex{\hspace{0em}f}%

\global\long\def\kpiccf{F}%

\global\long\def\kpiccfSet{\mathcal{F}}%

\global\long\def\floatingPointNumbers{\mathbb{F}}%

\global\long\def\sNatIndex{g}%

\global\long\def\kpiBorderRatio#1{g_{#1}}%

\global\long\def\specialkpiBorderRatio{g_{*}}%

\global\long\def\kpicca{G}%

\global\long\def\kpiBorderRatioSet{\mathcal{G}}%

\global\long\def\kpiBorderRatioInterval#1{\bar{\mathcal{G}}_{#1}}%

\global\long\def\kpiBorderConfigSet{\mathcal{H}}%

\global\long\def\kpiccb{H}%

\global\long\def\kpiBorderConfig{h}%

\global\long\def\kpiccc{I}%

\global\long\def\dataIndexSet{\mathcal{I}}%

\global\long\def\kpiccd{J}%

\global\long\def\orthogbasisIndexpairSet{\mathcal{J}}%

\global\long\def\innerNeighborhoodSize{\check{k}}%

\global\long\def\outerNeighborhoodSize{\hat{k}}%

%\global\long\def\knump{K}%

\global\long\def\numpapprox{K}%

\global\long\def\numptotalapprox{K_{\pwpolA}}%

\global\long\def\taylorApproxOrder{m}%

\global\long\def\kpiccIntegral{M}%

\global\long\def\mergeIndexSet{\mathcal{M}}%

\global\long\def\numApproxSamples{N}%

\global\long\def\bigO#1{\fc O{#1}}%

\global\long\def\pa{\mathbf{p}}%

\global\long\def\kpi{P}%

\global\long\def\kpipart#1#2{\kpi_{#1#2}}%

\global\long\def\kpipartup#1{\hat{\kpi}_{#1}}%

\global\long\def\kpipartdown#1{\check{\kpi}_{#1}}%

\global\long\def\kpipartborder#1{\bar{\kpi}_{#1}}%

\global\long\def\kpicomp#1#2#3#4{p_{#1#2#3#4}}%

\global\long\def\kpicompup#1#2#3#4#5{\hat{p}_{#1#2#3#4,#5}}%

\global\long\def\kpicompdown#1#2#3#4#5{\check{p}_{#1#2#3#4,#5}}%

\global\long\def\kpicompborder#1#2#3{\bar{p}_{#1#2#3}}%

\global\long\def\kSubBorder{q}%

\global\long\def\searchAreaMinRadiusRatio{Q}%

\global\long\def\positionalQE{Q_{\text{pos}}}%

\global\long\def\coefficientsQE#1{Q_{\text{val }#1}}%

\global\long\def\QE{Q_{\degapprox}}%

\global\long\def\bestQE{\QE^{*}}%
\global\long\def\bestQEsquared{\QE^{*\;2}}%

\global\long\def\kernelArgument{r}

\global\long\def\searchAreaInnerRadius{R}%

\global\long\def\rInterval#1{\mathcal{R}_{#1}}%

\global\long\def\searchAreaOuterRadius{S}%

\global\long\def\approxKsctSet{\mathcal{S}}%

\global\long\def\approxKsct{S}%

\global\long\def\bestApproxKsct{\approxKsct_{\adst}}%

\global\long\def\approxKsctCoeff{s}%

\global\long\def\sIntVar{s}%

\global\long\def\mortonBoxSize{s}%

\global\long\def\atomicStateVariable{s}%

\global\long\def\numTimeSteps{T}%

\global\long\def\mortonBoxInterval{\mathcal{B}}%

\global\long\def\taylorApprox#1{T_{#1}}%

\global\long\def\tInterval{\mathcal{T}}%

\global\long\def\knotPosition{t}%

\global\long\def\mortonBoxIntervalBorder{t}%

\global\long\def\tIntVar{t}%

\global\long\def\numScalar{U}%

\global\long\def\kpiPresubdomain{\mathcal{U}}%

\global\long\def\kpiSubdomain{\mathcal{U}}%

\global\long\def\kpiPresubdomainSet{\tilde{\mathscr{\mathfrak{U}}}}%

\global\long\def\kpiSubdomainSet{\mathscr{\mathfrak{U}}}%

\global\long\def\numVector{V}%

\global\long\def\ks{w}%

\global\long\def\ksk#1{\ks_{#1}}%

\global\long\def\ksc#1#2{\ksk{#1#2}}%

\global\long\def\pw{W}%

\global\long\def\approxCandidate{A}%

\global\long\def\approxCandPiece#1{A_{#1}}%

\global\long\def\approxFromLUT{\Tilde{A}}%

\global\long\def\approxCandCoeff{a}%

\global\long\def\kpiBorderFunctionsSet{\mathcal{Z}}%

\global\long\def\kpiSubSubdomain#1#2{\bar{\mathcal{Z}}_{#1#2}}%

\global\long\def\Zvalue{Z}%

\global\long\def\scalarField{\alpha}%

\global\long\def\vf{\boldsymbol{\beta}}%

\global\long\def\vfc{\beta}%

\global\long\def\objectiveField{\gamma}%

\global\long\def\objectiveFieldMultiplier#1{\indexedField{\hat{\objectiveField}}{#1}}%

\global\long\def\dst#1#2{\fc{\delta}{#1,#2}}%

\global\long\def\maxdst#1#2{\fc{\delta_{\max}}{#1,#2}}%

\global\long\def\mindst#1#2{\fc{\delta_{\min}}{#1,#2}}%

\global\long\def\pdst{\Delta}%

\global\long\def\pl{\zeta}%

\global\long\def\multipleContributionsFactor{\eta}%

\global\long\def\stirlingAbbreviation{\eta_{de}}%

\global\long\def\normalizedKnotPosition{\theta}%

\global\long\def\intVarTheta{\theta}%

\global\long\def\kpicctheta{\Theta}%

\global\long\def\stirlingCoeffA#1#2{\vartheta_{#1#2}}%

\global\long\def\knorm{\kappa}%

\global\long\def\knormSpecial{\kappa'}%

\global\long\def\kconstm{\kappa_{\taylorApproxOrder}'}%

\global\long\def\kconst{\tilde{\kappa}}%

\global\long\def\bitlength{\lambda}%

\global\long\def\kpicoeff{\lambda_{\kpiSubdomain abc}}%

\global\long\def\laurentCoeff{\lambda}%

\global\long\def\adst{\Lambda}%

\global\long\def\normadst{\adst}%

\global\long\def\pma{\mu}%

\global\long\def\stirlingCoeffC#1#2#3{\xi_{#1#2#3}}%

\global\long\def\sNatAbbr{\varpi_{de}}%

\global\long\def\pd{\rho}%

\global\long\def\searchAreaRadiiRatio{\rho_{*}}%

\global\long\def\stirlingCoeffB#1#2#3{\varrho_{#1#2#3}}%

\global\long\def\kpiccsign{\sigma}%

\global\long\def\coefficientsQV{\varsigma}%

\global\long\def\coefficientsQVfactor#1{\bar{\varsigma}_{#1}}%

\global\long\def\intVarTau{\tau}%

\global\long\def\hierarchyLevelDistance{\tau}%

\global\long\def\positionalQV{\tau}%

\global\long\def\bestPositionalQV{\positionalQV_{\degapprox}^{*}}%

\global\long\def\fallfacdiff#1#2#3{\fc{\varUpsilon_{#3}^{#2}}{#1}}%

\global\long\def\pc{\varphi}%

\global\long\def\pcmax{\phi}%

\global\long\def\pcrepr{\representative{\phi}}%

\global\long\def\stirlingCoeffD#1#2#3{\phi_{#1#2#3}}%

\global\long\def\kpiccphi{\Phi}%

\global\long\def\po{\boldsymbol{\chi}}%

\global\long\def\attributeOptAbbreviation{\psi_{ij}}%

\global\long\def\attributeOptAbbreviationA{\bar{\psi}_{i}}%

\global\long\def\kpiccpsi{\Psi}%

\global\long\def\kpiHalfSpace{\Omega}%

\global\long\def\kpiBorderPlane{\bar{\Omega}}%

\section{\uppercase{Introduction}}
\label{sec:introduction}

Introduced by Gingold and Monaghan \cite{Gingold1977} and independently by
Lucy \cite{Lucy1977}, \textbf{Smoothed Particle Hydrodynamics} (SPH) is a group of methods for simulating dynamic mechanical processes, typically fluid or gas
flows but also solid mechanics. The objective matter is modeled by means of particles, each representing a small portion of a simulated substance and attributed a specific mass, density, and other physical measures.

These discrete and scattered particles define scalar and vector fields on the spatial continuum through an interpolation rule, determining a physical field as the sum of the isotropic contributions of the particles, each following a smooth function only of the distance
from the particle position, called the kernel function.
As an example, we refer to the SPH kernel function defined by the cubic B-spline \cite{Rosswog2009}
\begin{equation}
    \fc{\ks}{\kernelArgument}=\frac{1}{4\pi}\begin{cases}
\left(2-\kernelArgument\right)^{3}-4\left(1-\kernelArgument\right)^{3}, & 0\le\kernelArgument<1\\
\left(2-\kernelArgument\right)^{3}, & 1\le\kernelArgument<2\\
0, & 2\le\kernelArgument\;.
\end{cases}
\label{eq:cubic-kernel}
\end{equation}
In this work, we assume the kernel function to have compact support. We consider this to be a minor restriction because kernel functions with compact support are appreciated in the SPH community for bounding the particles' volumes of influence. Obviously, by simply defining some cut-off value as upper bound, any kernel function can be supported.
Given such a function, a particle's contribution to a scalar field can be expressed as
\[\frac{\pma\scalarField}{\pd\pl^{3}}\fc{\ks}{\frac{\left\Vert \v x-\po\right\Vert }{\pl}}\]
for the particle's position \(\po\), mass \(\pma\), target field value \(\scalarField\), density \(\pd\), and smoothing radius \(\pl\). While \(\pma\), \(\scalarField\), and \(\pd\) serve as simple multiplicative constants, defining only the amplitude of the contribution, \(\pl\) radially scales the
domain and thus defines the radius of the particle’s volume of influence. 

\textbf{Direct Volume Rendering (DVR)} has a long-standing tradition for visualizing scalar volumetric fields \cite{Drebin1988} and is commonly implemented as a ray casting method. It builds on assigning visual characteristics, representing features
of the field, to the target domain volume and then rendering it by casting viewing rays through the volume. The color of each pixel of the output image is the result of simulating the behavior of light traveling through
the volume along a ray in opposite viewing direction.

Said visual characteristics, usually comprising light emission and absorption, are computed from local target field characteristics, like values or gradients, in a process called classification \cite{Max1995}. It is commonly employed based on a piecewise constant or piecewise linear approximation of the target scalar field along the ray.

Direct volume rendering of astronomical SPH data was first performed by   
incorporating contributions of both grid and particle data into the optical
model \cite{Kahler2007}.
Efficient full-featured DVR applied directly to large SPH data sets
on commodity PC hardware has first been managed by resampling the scalar fields to a perspective
grid held in a \threed/ texture for each rendering frame 
\cite{Fraedrich2010}.
Later, DVR for large scattered data was proposed to be performed on the CPU by Knoll
\etal\cite{Knoll2014} and mostly applied to molecular dynamics. Although employing radial basis function kernel (RBF) interpolation similar to the SPH case, they focus on the very special task of rendering a surface defined by a density field. This is in line with more recent contributions in rendering SPH data, such as \cite{hochstetter2016adaptive} or \cite{oliveira2022narrow}, targeting only simulated fluids.

The approaches for volume-rendering arbitrary SPH fields employ equidistant sampling of the target field along viewing rays and acting on a piecewise linear approximation of it, which, depending on the local particle density, may miss detailed features in regions of high particle density, or oversample particles with large smoothing length in sparse regions.

Thus, in this work, we explore the capabilities and limitations of approximating SPH scalar fields with higher-order piecewise polynomial functions, whose resolution adapts
to the local resolution of the given particle data. The higher-order approximations may facilitate quantitatively more
accurate output images at a worthwhile cost.

\section{\uppercase{Method Overview}}

As SPH interpolation defines the target scalar field as the sum of particle contributions to it, our concept builds on approximating each individual particle contribution on a viewing ray and then employing the sum of these approximations. Although summing an indefinite number of piecewise polynomial functions may seem like a task of quadratic complexity not amenable to GPU processing, we encode these functions in a way reducing the summation to a simple sorting task.

Our method comprises three passes:
\begin{enumerate}
        \item For each particle, approximate its contribution to all relevant viewing rays.
        \item For each ray, sort the contribution data assigned to it with respect to the distance from the viewer.
        \item For each ray, accumulate the contributions along the ray, classify and composite the result.
\end{enumerate}

In the remainder of this section, we declare our concept of representing piecewise polynomial functions and show how these can be processed during the compositing sweep. In \Section \ref{sec:approximating-single-particle-contributions}, we present a way to efficiently compute optimal approximations for single particle contributions during the first sweep. In \Section \ref{sec:quantization-to-prevent-higher-order-errors}, we then describe a pitfall that our higher-order approximation scheme involves and develop an improvement of our method to overcome this difficulty.
Finally, in \Section \ref{sec:choosing-approximation-dimensions}, we analyze the approximation errors implied by our scheme and discuss the choice of the major parameters like the approximation order, before concluding our work in \Section \ref{sec:conclusion}.

\paragraph{Localized Difference Coefficients.}
A real polynomial function $\pwpolA: \mathbb{R}\to\mathbb{R}$ of order $\degapprox$ is generally represented by its coefficients, \ie numbers $\pwpolACoeff_d\in\mathbb R$ such that
\(\fc{\pwpolA}{\knotPosition}=\sum_{d=0}^{\degapprox}\pwpolACoeff_{d}\knotPosition^d\). They provide a direct image of how the function and all its derivatives behave at $\knotPosition=0$ since the $d^\text{th}$ derivative of $\pwpolA$ at $\knotPosition=0$ amounts to $d!\pwpolACoeff_d$. If we wanted to know the value or derivative of order $d$ at some other $\knotPosition_*$, we could compute it as
\[
\fc{\pwpolA^{\left(d\right)}}{\knotPosition_*}=\sum_{j=d}^{\degapprox}\frac{j!}{\left(j-d\right)!}\pwpolACoeff_{j}\knotPosition_*^{j-d}
=d!\sum_{j=d}^{\degapprox}\binom{j}{d}\pwpolACoeff_{j}\knotPosition_*^{j-d}
,\]
which shows that the real numbers \[\pwpolACoeff_{*d}=\sum_{j=d}^{\degapprox}\binom{j}{d}\pwpolACoeff_{j}\knotPosition_*^{j-d}\] represent the order-$d$ behavior of $\pwpolA$ at $\knotPosition=\knotPosition_*$ just as the $\pwpolACoeff_d$ do at $\knotPosition=0$. In fact, $\pwpolA$ can be expressed using these \textit{localized coefficients} as 
\[\fc{\pwpolA}{\knotPosition}=\sum_{d=0}^{\degapprox}\pwpolACoeff_{*d}\left(\knotPosition-\knotPosition_{*}\right)^d.\]

Now, a continuous piecewise polynomial function is defined by a sequence of border arguments $\knotPosition_0 < \knotPosition_1 < \dots$ and several polynomial functions $\pwpolA_0, \pwpolA_1, \dots$ such that each polynomial function $\pwpolA_k$ is applied in its respective interval $\left[\knotPosition_{k},\knotPosition_{k+1}\right]$.

If we were representing each polynomial $\pwpolA_k$ by its coefficients, summing several piecewise polynomial functions would require tracking the applicable polynomials for each resulting argument interval and summing their coefficients.

Instead, we save for each border argument $\knotPosition_k$ the change that the overall function performs in all orders. Specifically, we save the localized coefficients of the difference between the polynomials $\pwpolA_k$ applicable on the right of $\knotPosition_k$ and $\pwpolA_{k-1}$ applicable on its left.

Thus, when saving a piecewise polynomial approximation of order \(\degapprox\) along a viewing ray, we encode it as a sequence of \textit{knots}
\[\left(\knotPosition_{k},\hat{\pwpolACoeff}_{k0},\dots,\hat{\pwpolACoeff}_{k\degapprox}\right),\quad k = 0, 1, \dots \]
consisting of a ray parameter \(\knotPosition_k\) defining the knot position on the ray and \textit{localized difference coefficients} \(\hat{\pwpolACoeff}_{k0},\dots,\hat{\pwpolACoeff}_{k\degapprox}\), such that
\[\fc{\pwpolAPiece k}{\knotPosition}-\fc{\pwpolAPiece{k-1}}{\knotPosition}=\sum_{d=0}^{\degapprox}\hat{\pwpolACoeff}_{kd}\left(\knotPosition-\knotPosition_{k}\right)^d.\]
Summing several piecewise polynomials encoded this way amounts to nothing more than sorting their joint knots for increasing knot positions. The resulting approximation can then be processed piece by piece, retrieving the localized coefficients 
$\pwpolACoeff_{k0},\dots,\pwpolACoeff_{k\degapprox}$ of each piece polynomial $\pwpolA_k$ at its left boundary argument $\knotPosition_k$ from the ones of the last piece according to the updating rule
\begin{equation}\label{eq:Piecewise-Coefficients-Update}\pwpolACoeff_{kd}=\hat{\pwpolACoeff}_{kd}+\sum_{j=d}^{\degapprox}\binom{j}{d}\pwpolACoeff_{\left(k-1\right)j}\left(\knotPosition_{k}-\knotPosition_{k-1}\right)^{j-d}\end{equation}
for \(d=0,\dots,\degapprox\), which is a computation of constant complexity per piece, irrespective of the number of particles contributing.

\section{\uppercase{Approximating Single Particle Contributions}\label{sec:approximating-single-particle-contributions}}

\subsection{Deduction from Unit Particle\label{subsec:deduction-from-unit-particle}}
Since the same SPH kernel function is applied for all particles, their contributions to the volume differ in only a few scaling and translation parameters, namely their position \(\po\), mass \(\pma\), density \(\pd\), smoothing radius \(\pl\), and applicable scalar field attribute \(\scalarField\).
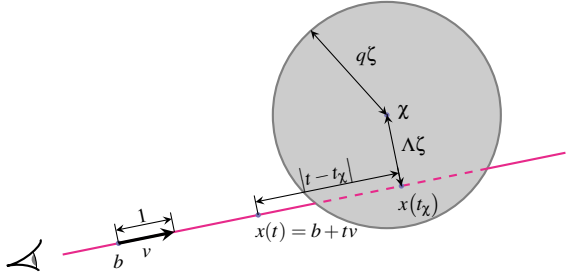
\begin{figure}[H]
  \centering
  \newcommand{\eye}[2][0]%position,rotation
{
	\begin{scope}[shift={(#2)},rotate=#1]
\draw (0,0)..controls+(0.5cm,0)and+(0.1cm,-0.1cm)..+(1cm,0.5cm)
	 +(0,0)..controls+(0.5cm,0)and+(-0.0,0.05cm)..
	 +(0.8cm,-0.3cm);
\fill[gray] (0,0)+(-12:0.78cm) arc[radius=0.78cm, start angle=-12, end angle=12] arc [radius=0.187259078cm,start angle=120, end angle=240];
\fill[black] (0,0)+(-3:0.78cm) arc[radius=0.78cm, start angle=-3, end angle=3] arc [radius=0.063507829cm,start angle=140, end angle=220];
\draw[thin] (0,0)+(-17.275:0.78cm) arc[radius=0.78cm, start angle=-17.275, end angle=16.77];
\end{scope}
}
\newcommand{\distance}[3][5pt]%start, stop,distance
{
\begin{scope}
\coordinate (startDistance__) at (#2);
\coordinate (endDistance__) at (#3);
\coordinate (arrowStart__) at ($(startDistance__)!#1!90:(endDistance__)$);
\coordinate (arrowEnd__) at ($(endDistance__)!#1!270:(startDistance__)$);
\draw[thin] (startDistance__) -- ($(startDistance__)!1.5!(arrowStart__)$);
\draw[thin] (endDistance__) -- ($(endDistance__)!1.5!(arrowEnd__)$);
\draw[<->,>=stealth,thin] (arrowStart__) -- (arrowEnd__);
\end{scope}
} 

\centering
\begin{tikzpicture}[thick, scale=0.5, every node/.style={scale=0.75}]
\tikzmath{\rot=asin(3/14.7);}
\filldraw[gray,fill=black!20,name path=part] (0,0) circle (3cm);
\coordinate (eyeback) at (-10,-4);
\coordinate (rayend) at (4.7,-1);
\coordinate (a) at ($(eyeback)!0.2!(rayend)$);
\eye[\rot]{eyeback}
\path[name path=ray] (eyeback) -- (rayend);
\path[name intersections={of=part and ray}];
\draw[myred]($(eyeback)!0.1!(rayend)$) -- (intersection-1);
\draw[myred,dashed,dash phase=3pt](intersection-1) --(intersection-2);
\draw[myred] (intersection-2) -- (rayend);
\fill[myblue] (0,0) circle (2pt);
\coordinate (nearest) at ($(intersection-1)!0.5!(intersection-2)$);
\fill[myblue] (a) circle (2pt);
\coordinate (apv) at ($(eyeback)!0.3!(rayend)$);
\draw[->,>=stealth,very thick] (a) -- (apv);
\node[anchor=base west,xshift=2pt] at (0,0) {$\po$};
\node[anchor=north,inner sep=5pt] at (a) {$\rayBasePoint$};
\node[anchor=north, inner sep=5pt] at ($(eyeback)!0.25!(rayend)$) {$\v v$};
\distance[10pt]{a}{apv}
\coordinate(aphv) at ($(a)!0.5!(apv)$);
\node[rotate=\rot] at ($(aphv)!0.6cm!90:(apv)$) {1};
\fill[myblue] (nearest) circle (2pt);
\node[anchor=north west,xshift=-5pt,yshift=-0pt] at (nearest) {$\fc{\v x}{\knotPosition_{\po}}$};
\distance[0pt]{$(0,0)!0pt!(nearest)$}{$(nearest)!0pt!(0,0)$}
\node[anchor=base west,xshift=0pt] at ($(0,0)!0.5!(nearest)$) {$\adst\pl$};
\coordinate (radius) at (-2,{sqrt(5)});
\distance[0pt]{radius}{$(0,0)!0pt!(radius)$}
\node[anchor=south west,xshift=0pt,yshift=0pt] at ($(0,0)!0.5!(radius)$) {$\kSubBorder\pl$};
\coordinate (apoint) at ($(eyeback)!0.45!(rayend)$);
\fill[myblue] (apoint) circle (2pt) ;
\node[anchor=north west,xshift=-5pt,yshift=-0pt] at (apoint) {$\fc{\v x}{\knotPosition}=\rayBasePoint+\knotPosition \v v$};
\distance[10pt]{apoint}{nearest}
\node[rotate=\rot] at (-1.6,-1.6) {$\left|\knotPosition-\knotPosition_{\po}\right|$};
\coordinate (rightanglestart) at ($(nearest)!0.3cm!(0,0)$);
\coordinate (rightangleend) at ($(nearest)!-0.3cm!(eyeback)$);
\end{tikzpicture}
  \caption{Sketch of measures involved in the positional relationship between viewing ray and particle. The volume of influence of a particle with smoothing length \(\pl\) is intersected by a viewing ray, defined by base point \(\rayBasePoint\) and unit direction vector \(\v v\). The particle's contribution on the ray at a point \(\fc{\v x}{\knotPosition}\) is determined by its distance to the particle position \(\po\). We denote by \(\kSubBorder\) the upper bound of the kernel function's support, such that \(\kSubBorder\pl\) is the radius of the particle's volume of influence.}
  \label{fig:ray-to-particle-position}
 \end{figure}

To model a viewing ray, we fix a straight line with vector equation \(\fc{\v x}{\knotPosition}=\rayBasePoint+\knotPosition\v v\) for some base point \(\rayBasePoint\in\mathbb{R}^{3}\), unit direction vector \(\v v\in\mathbb{R}^{3}\), and parameter \(\knotPosition\in\mathbb{R}\).
We consider a normalized particle, \ie one of unit mass, unit density, unit field attribute, and unit smoothing radius. Assuming the base point \(\rayBasePoint\) on the ray is the one closest to the particle position, the normalized particle's contribution to this line is
\[\fc{\ksct{\adst}}{\knotPosition}=\fc w{\sqrt{\adst^{2}+\knotPosition^{2}}},\]
where \(\adst\) is the distance between the particle's position and the line.
Then, the contribution of a specific data particle on a ray at distance \(\adst\pl\) from its position \(\po\) amounts to
\[\frac{\pma\scalarField}{\pd\pl^{3}}\fc{\ksct{\adst}}{\frac{\knotPosition-\knotPosition_{\po}}{\pl}},\]
where \(\knotPosition_{\po}\) is the parameter of the point on the ray closest to the particle position and \(\adst=\frac{1}{\pl}\left\Vert\po-\fc{\v x}{\knotPosition_{\po}}\right\Vert\). The situation is depicted in \Figure \ref{fig:ray-to-particle-position}.

Finding optimal piecewise polynomial approximations for \(\ksct{\adst}\)  for all \(\adst\) within the SPH kernel's support suffices to generate optimal approximations for any particle by just translating and scaling it in the same way. 
For quick access, we thus prepare a look-up table containing the localized difference coefficients of order-\(\degapprox\) approximations of \(\ksct{\adst}\) for many equidistant values of \(\adst\). In the remainder of this section, we show how we can find these optimal approximations.

\subsection{Optimization Problem Definition\label{subsec:optimization-problem-definition}}
Before we can find optimal approximations, we need to define what optimality shall mean in this context. Specifically, we have to settle on:
\begin{enumerate}
    \item The space of eligible candidates, \ie the condition of what we want to consider a feasible approximation.
    \item The error measure defining whether one approximation is better than another.
\end{enumerate}
The main restriction on the space of eligible approximation candidates is the imposition of a maximum polynomial degree \(\degapprox\), i.e, the order of approximation, and the number \(\numpapprox\) of non-trivial polynomial pieces per particle, \ie the number of non-zero polynomials defining the approximation of a single particle. We dedicate \Section \ref{sec:choosing-approximation-dimensions} to evaluating choices of \(\degapprox\) and \(\numpapprox\) but leave them unspecified for now. Beyond that, we demand that our approximations shall be continuous and even functions with compact support. This seems reasonable, as our approximation target \(\ksct{\adst}\) also has these properties.

For measuring the approximating quality of any piecewise polynomial function candidate \(\approxKsct:\mathbb{R}\to\mathbb{R}\), we apply its \(L^{2}\) distance to \(\ksct{\adst}\), \ie we seek to minimize the approximation error 
\begin{equation}
\fc{\approxError}{\approxKsct}=\left\Vert \approxKsct-\ksct{\adst}\right\Vert _{2}=\left(\intop_{\mathbb{R}}\left[\fc{\approxKsct}{\knotPosition}-\fc{\ksct{\adst}}{\knotPosition}\right]^{2}\diff\knotPosition\right)^{\frac{1}{2}}\!\!\!.
\label{eq:distance-related-approximation-error}
\end{equation}
In contrast to the supremum norm, often used in approximation theory, this error measure punishes not only the maximum pointwise deviation from the target, but also the length of segments of high deviation. Moreover, another advantage  of the \(L^{2}\) norm is its associated inner product, which allows us to generate a closed-form solution of the optimal approximation and its error value as a function only of the knot positions, as shown in \Section \ref{subsec:solution-for-fixed-knot-positions}. We then find optimal knot positions through standard non-linear optimization as explained in \Section \ref{subsec:optimal-knot-positions}.

\subsection{Solution for Fixed Knot Positions\label{subsec:solution-for-fixed-knot-positions}}
As a prerequisite for finding truly optimal approximations, we first handle the case of arbitrary fixed knot positions. Due to our evenness requirement, the negative knot positions are determined from the positive ones. Hence, our optimization domain is the set of even continuous piecewise polynomial functions with compact support, maximal degree \(\degapprox\), maximal number of non-trivial pieces \(\numpapprox\), and positive knot positions \(\normalizedKnotPosition_{1},\dots,\normalizedKnotPosition_{\lastKnotIndex}\). We denote this set by \(\approxKsctSet\).

Consider the vector space \(\fc{L^{2}}{\mathbb{R}}\) of square-integrable functions \(\mathbb{R}\to\mathbb{R}\), which by the Riesz-Fischer theorem is complete with respect to the \(L^{2}\) norm and therefore a Hilbert space when equipped with the inner product
\[\left\langle \squareIntegrableA,\squareIntegrableB\right\rangle =\intop_{\mathbb{R}}\fc{\squareIntegrableA}{\knotPosition}\fc{\squareIntegrableB}{\knotPosition}\diff\knotPosition\quad\text{for}\quad\squareIntegrableA,\squareIntegrableB\in\fc{L^{2}}{\mathbb{R}},\]
which induces the \(L^{2}\) norm \(\left\Vert \squareIntegrableA\right\Vert _{2}=\sqrt{\left\langle \squareIntegrableA,\squareIntegrableA\right\rangle }\). The approximation target \(\ksct{\adst}\) is clearly an element of \(\fc{L^{2}}{\mathbb{R}}\) as it is continuous and has compact support.

For any \(\squareIntegrableA,\squareIntegrableB\in\fc{L^{2}}{\mathbb{R}}\), \(\squareIntegrableB\not\equiv\mathbf{0}\), we denote by \[\proj{\squareIntegrableB}{\squareIntegrableA}=\frac{\left\langle \squareIntegrableA,\squareIntegrableB\right\rangle }{\left\langle \squareIntegrableB,\squareIntegrableB\right\rangle }\squareIntegrableB\]
the orthogonal projection of \(\squareIntegrableA\) on \(\squareIntegrableB\).

As we will see shortly, \(\approxKsctSet\) is a subspace of \(\fc{L^{2}}{\mathbb{R}}\) of finite dimension \(\left\lfloor \frac{\numpapprox\degapprox}{2}\right\rfloor \), for which we can compute an orthogonal basis \(\orthogbasis\), which in turn we can use to calculate the orthogonal projection of \(\ksct{\adst}\) on \(\approxKsctSet\) as
\begin{equation}
\label{eq:optimal-approx-for-given-knots}
\bestApproxKsct=\sum_{\pwpolA\in\orthogbasis}\proj{\pwpolA}{\ksct{\adst}}.
\end{equation}

It is easy to show that \(\bestApproxKsct\) is the unique error-optimal approximation of \(\ksct{\adst}\) among the elements of \(\approxKsctSet\) (proof in Appendix, \Section \ref{appsec:optimality}).
Therefore, all we need for computing the optimal approximation for fixed knot positions is a suitable orthogonal basis.

We specify a non-orthogonal basis as a starting point here: Let \(\orthogbasisIndexpairSet\) be the set of pairs \(\left(k,d\right)\) of positive integers \(k\le\lastKnotIndex\) and \(d\le\degapprox\) but excluding elements \(\left(1,d\right)\) for uneven \(d\) if \(\numpapprox\) is uneven. Then the set \(\nonorthogbasis=\left\{ \vphantom{\big\{}\smash{\nonorthogbasiselement kd}:\left(k,d\right)\in\orthogbasisIndexpairSet\right\}\) of functions
\[\fc{\nonorthogbasiselement kd}{\knotPosition}=\begin{cases}
1 & \text{if }\left|\knotPosition\right|\le\normalizedKnotPosition_{k-1}\\
1-\left(\frac{\left|\knotPosition\right|-\normalizedKnotPosition_{k-1}}{\normalizedKnotPosition_{k}-\normalizedKnotPosition_{k-1}}\right)^{d} & \text{if }\normalizedKnotPosition_{k-1}\leq\left|\knotPosition\right|\leq\normalizedKnotPosition_{k}\\
0 & \text{if }\left|\knotPosition\right|\ge\normalizedKnotPosition_{k}
\end{cases}\]
is a basis of \(\approxKsctSet\) (proof in Appendix, \Section \ref{appsec:basis}).

Given \(\nonorthogbasis\), we convert it into an orthogonal basis $\orthogbasis=\left\{ \orthogbasiselement kd:\left(k,d\right)\in\orthogbasisIndexpairSet\right\} $
by employing the Gram-Schmidt process. Specifically, we recursively
set
\[
\orthogbasiselement{k^{*}}{d^{*}}=\nonorthogbasiselement{d^{*}}{k^{*}}-\sum_{\substack{\left(k,d\right)\in\orthogbasisIndexpairSet\\
k<k^{*}\vee\left(k=k\wedge d<d^{*}\right)
}
}\proj{\orthogbasiselement kd}{\vphantom{\big(}\smash{\nonorthogbasiselement{d^{*}}{k^{*}}}}
\]
 in lexicographical order of pairs $\left(k^{*},d^{*}\right)\in\orthogbasisIndexpairSet$.

\subsection{Optimal Knot Positions\label{subsec:optimal-knot-positions}}
While we can directly compute optimal approximations for given knot
positions as shown above, finding error-optimal knot
positions is a nonlinear optimization problem over the variables $\normalizedKnotPosition_{1},\dots,\normalizedKnotPosition_{\lastKnotIndex}$.

The objective function $\bestApproxError{\adst}$ is continuous
within the interior of the feasible domain defined by the constraints
$0<\normalizedKnotPosition_{1}<\dots<\normalizedKnotPosition_{\lastKnotIndex}$,
due to the continuity of the inner product and of the constructed basis with respect to the \(\normalizedKnotPosition_k\),
which would even hold for a discontinuous SPH kernel function. Also, we can
expect to find a global optimizer within the interior of the feasible
domain, \ie without any of the constraints being active, because
an equality of any two variables would be equivalent to a reduction
of the number of non-zero pieces, diminishing the freedom for approximating
and therefore resulting in higher or equal error. Hence, if a local optimum
was attained at the feasible domain border, it could not be isolated
because shifting one of the colliding $\normalizedKnotPosition_{k}$
and defining the polynomials on both of its sides to be equal would result
in the same approximation function and therefore in the same error
value.

While evaluating $\bestApproxError{\adst}$ could be done following
the steps above
for every set of fixed $\normalizedKnotPosition_{1},\dots,\normalizedKnotPosition_{\lastKnotIndex}$,
it is worthwhile to only fix $\numpapprox$ and $\degapprox$ and
perform the process in a symbolic manner, generating an explicit formula
of the objective error function, which can later be evaluated for
any knot positions and distance parameter $\adst$. The generation
of a closed-form representation of an orthogonal basis for variable
knot positions and $\adst$ has to be done
only once as it does not depend on the SPH kernel used. However, the
explicit expression of the approximation error following (\ref{eq:optimal-approx-for-given-knots})
requires closed-form solutions of the integrals defining the inner
products $\left\langle \pwpolA,\ksct{\adst}\right\rangle $ for orthogonal
basis functions $\pwpolA$. In the case of an SPH kernel defined by
a continuous piecewise polynomial function such as the cubic B-spline
kernel (\ref{eq:cubic-kernel}), this can clearly be achieved. In any case, as the
symbolic computations are rather involved, computer algebra systems
are of great help during this preparatory process.

We have conducted this process for the cubic B-spline kernel, going to considerable length to find
a global optimizer for many discrete $\adst$. For $\adst$  close to the upper bound \(\kSubBorder\), however,
we have found the evaluation of some of the formulae to become instable.
To obtain reliable results, we employed the GNU MPFR library to perform the computations in multiple precision
arithmetic.
Since we have not encountered any severe problems during
these optimizations, we have reason to hope they are manageable
for any piecewise polynomial SPH kernel function.

\section{\uppercase{Quantization to Prevent Higher-order Errors}\label{sec:quantization-to-prevent-higher-order-errors}}

\begin{figure*}[t]
(a)~\raisebox{\dimexpr 1em-\totalheight\relax}{\includegraphics[width=7cm, trim={0 30cm 0 0}, clip]{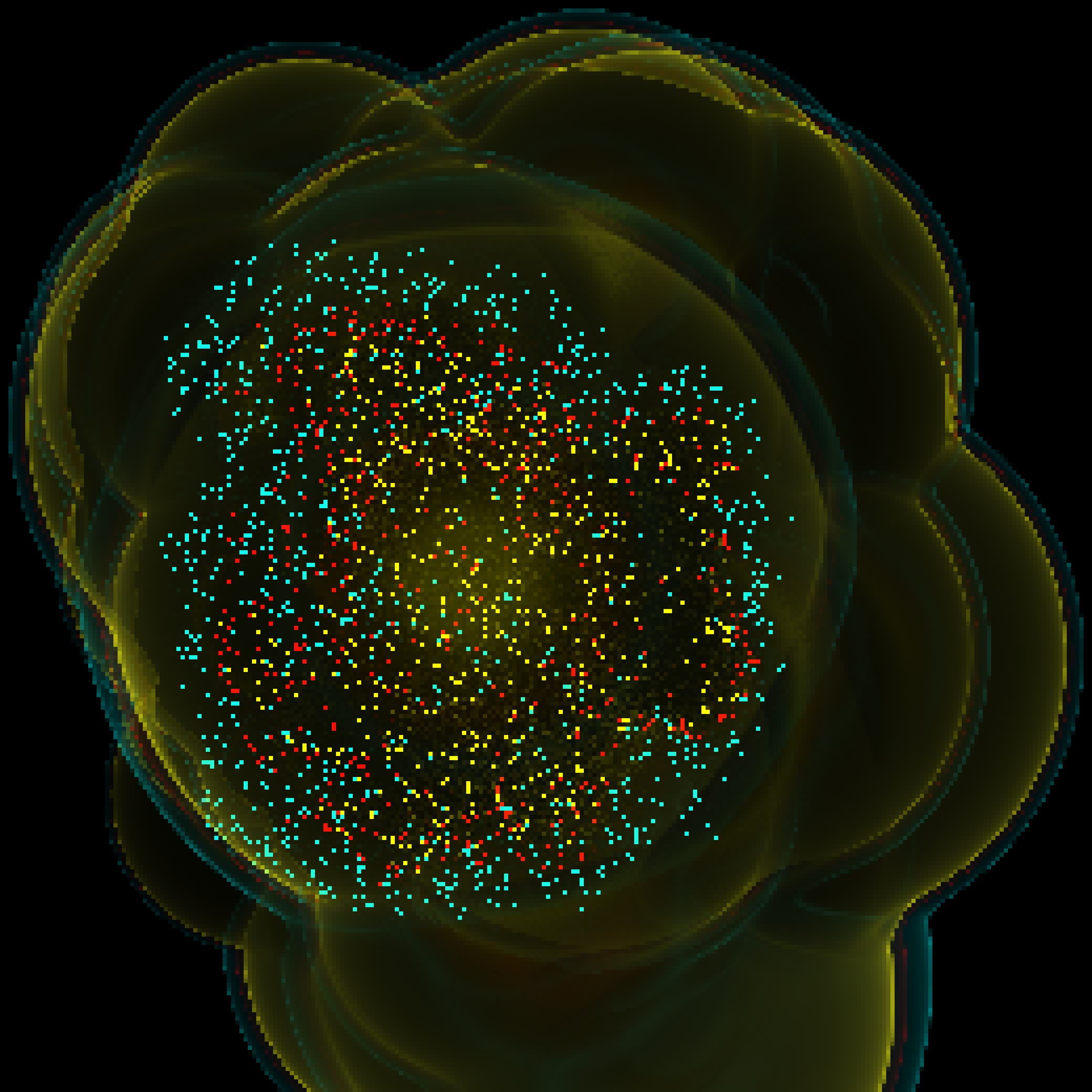}}\hspace*{\fill}(b)~\raisebox{\dimexpr 1em-\totalheight\relax}{\includegraphics[width=7cm, trim={0 30cm 0 0}, clip]{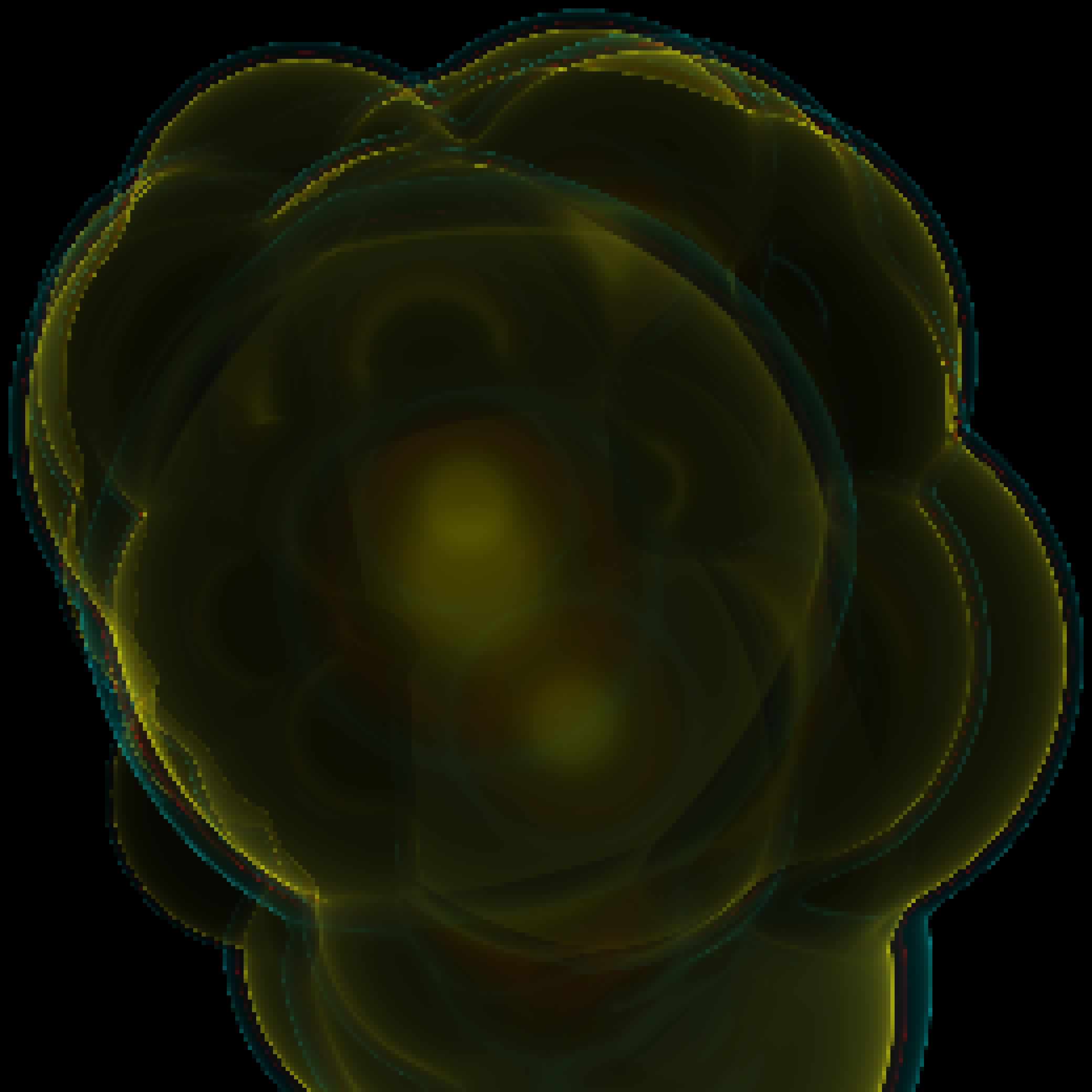}}

\caption[Effects of Higher-order Rounding Error Propagation]{\label{fig:sample-error-propagation-artifacts}
Extracts of sample renderings of the temperature field of an SPH data set using our higher-order SPH field approximation scheme and a transfer function emphasizing three rather low temperature value regions by mapping them to an emission of blue, yellow, and red light. All computations are performed in single precision on the GPU. \protect \\
(a) clearly shows ``sprinkling'' artifacts caused by higher-order rounding error propagation, which ``randomly'' cause the field approximation on the ray to stay within one of the highlighted temperature regions far ``behind'' the particle cluster. \protect \\
(b) shows the result of employing 10 slices of reinitialization to mitigate the problem.
}
\end{figure*}

\subsection{Higher-order Rounding Error Propagation}

For the stated efficiency reasons, the recursive computation of polynomial
coefficients according to (\ref{eq:Piecewise-Coefficients-Update})
is an integral component to our particle-wise approximation approach.
However, it comes at a substantial price which may not be obvious
at first glance. Directly applying (\ref{eq:Piecewise-Coefficients-Update}) during the compositing sweep
using floating-point numbers is problematic because rounding errors
are propagated at higher order from front to back along the viewing ray, resulting in unreliable
coefficients especially for higher $\knotPosition$.

To illustrate the issue, consider a piecewise polynomial
function modeling the contribution of a small number of particles.
Clearly, the last piece polynomial of this approximation should be the zero polynomial.
However at its starting knot, its value is most probably not computed
to be zero but some value close to zero, due to rounding errors
in the computation. Although this zeroth-order distortion may
be negligible by itself, 
small rounding errors in higher-order terms cause
large errors further down the ray.
We may thus find the polynomial's value to have grown far from zero for larger \(\knotPosition\).

The direct effect of these errors is an unstable result: When using a transfer function with
focus on lower attribute values typically reached just before leaving
the volumes of influence of the last contributing particles, the resulting
pixel colors are highly unstable and the generated images show strong
“sprinkling” artifacts, as shown in \Figure \ref{fig:sample-error-propagation-artifacts}~(a). Hence, solving this problem is indispensable
if we want our higher-order ray casting concept to be of any use.

There are several conceivable measures for alleviation.
One is to limit the maximum distance on the ray that the error may
use to grow by introducing a number of special reinitialization knots
at predetermined \(\knotPosition\) on all rays, as performed for the rendering of \Figure\ref{fig:sample-error-propagation-artifacts}~(b). When processing a particle during the first sweep,
in addition to the knots encoding its higher-order approximation as difference coefficients,
we also add to all covered reinitialization knots the localized coefficients
of this particle's contributions. Later when processing the sorted
sequence of knots, whenever we encounter a reinitialization
knot, we directly take the coefficients attached to it instead of
computing the coefficients following the update rule (\ref{eq:Piecewise-Coefficients-Update}),
thus eliminating the effect of past errors for future pieces. However,
this approach not only introduces the complexity of two different
kinds of knots but
also can only partially solve the problem. Besides, defining how many reinitialization
knots to utilize is non-trivial.

Clearly, an alternative would be to avoid the cause of higher-order
error propagation altogether and proceed to a direct, possibly localized,
coefficient representation of the polynomials, considering for the
computation of each result piece only the contributions with overlapping
support. While this would most probably facilitate stable results,
it would mean accepting the expense of determining for each
piece the set of relevant contributions.

\subsection{Exact Arithmetic Through Quantization\label{subsec:exact-arithmetic-through-quantization}}

We propose yet another approach to fully avoid
rounding errors during the computation of (\ref{eq:Piecewise-Coefficients-Update}), namely by transferring all involved
quantities from floating-point to fixed-point numbers, which allow
an exact arithmetic.
More precisely, we slightly shift all numbers encoding the individual
contribution approximations into integer multiples of some quantum
values, a process we call quantization. While this increases the approximation
errors for the individual contributions, the update operations in
(\ref{eq:Piecewise-Coefficients-Update}) are reduced to exact integer
manipulations.

Adhering to the notation used in (\ref{eq:Piecewise-Coefficients-Update}),
the values to be quantized are the elements
of the knots $\left(\knotPosition_{k},\hat{\pwpolACoeff}_{k0},\dots,\hat{\pwpolACoeff}_{k\degapprox}\right)$
and the localized coefficients $\pwpolACoeff_{kd}$.
We thus seek to fix real quantum values $\positionalQV>0$
and $\coefficientsQV_d>0$
for $d=0,\dots,\degapprox$ such that
$
\knotPosition_{k}\approx\bar{\knotPosition}_{k}\positionalQV$ for some $\bar{\knotPosition}_{k}\in\mathbb{N}$
and
$\hat{\pwpolACoeff}_{kd}\approx\bar{\hat{\pwpolACoeff}}_{kd}\coefficientsQV_d$, $\pwpolACoeff_{kd}\approx\bar{\pwpolACoeff}_{kd}\coefficientsQV_d$ for integers $\bar{\hat{\pwpolACoeff}}_{kd}$ and $\bar{\pwpolACoeff}_{kd}$.
The quantized knots can then be encoded as the integer components
$\left(\bar{\knotPosition}_{k},\bar{\hat{\pwpolACoeff}}_{k0},\dots,\bar{\hat{\pwpolACoeff}}_{k\degapprox}\right)$.

Written in these terms, (\ref{eq:Piecewise-Coefficients-Update})
becomes
\begin{equation*}
\bar{\pwpolACoeff}_{kd}=\bar{\hat{\pwpolACoeff}}_{kd}+\sum_{j=d}^{\degapprox}\binom{j}{d}\frac{\coefficientsQV_j\positionalQV^{j-d}}{\coefficientsQV_d}\bar{\pwpolACoeff}_{\left(k-1\right)j}\left(\bar{\knotPosition}_{k}-\bar{\knotPosition}_{k-1}\right)^{j-d},
\end{equation*}
 which we have to guarantee to result in an integer for arbitrary
integer values of previously computed $\left(\bar{\knotPosition}_{k}-\bar{\knotPosition}_{k-1}\right)$
and $\bar{\pwpolACoeff}_{\left(k-1\right)j}$. This can only be accomplished
by requiring 
$\binom{j}{d}\frac{\coefficientsQV_j\positionalQV^{j-d}}{\coefficientsQV_d}\in\mathbb{Z}$ for all $d$ and $j\geq d$, which we can fulfill by simply setting $\coefficientsQV_d=\frac{\coefficientsQV}{\positionalQV^d}$ for all $d$, where we abbreviate $\coefficientsQV=\coefficientsQV_0$. This leaves us with only two quantum values: one length quantum \(\positionalQV\) and one field value quantum \(\coefficientsQV\). We show our strategy of choosing the two in \Section \ref{subsec:specifying-quantum-values}.

However, we first specify in \Section \ref{subsec:setting-quantized-knots} how we set up the quantized knot positions \(\bar{\knotPosition}_k\) and difference coefficients \(\bar{\hat{\pwpolACoeff}}_{kd}\), which later form the input of the coefficients update rule (\ref{eq:Piecewise-Coefficients-Update}), now reduced to the integer-only representation
\begin{equation}
\bar{\pwpolACoeff}_{kd}=\bar{\hat{\pwpolACoeff}}_{kd}+\sum_{j=d}^{\degapprox}\binom{j}{d}\bar{\pwpolACoeff}_{\left(k-1\right)j}\left(\bar{\knotPosition}_{k}-\bar{\knotPosition}_{k-1}\right)^{j-d}.\label{eq:accumulation-rule}
\end{equation}

\subsection{Setting Quantized Knots\label{subsec:setting-quantized-knots}}
In order to not introduce higher-order errors through the
back door after all, we have to ensure that the input to the integer computations does not contain such errors already. Each individual particle approximation has to have compact support, \ie its knots have to exactly neutralize each other.
This guides us to compute the knots of single particle contributions according to the following rules (derivation in Appendix, \Section \ref{appsec:quantized-knots}).

Given a particle with attributes $\pma$, $\pd$, $\pl$,  and $\scalarField$,  positioned at distance $\adst\pl$ from the ray with closest point parameter $\knotPosition_{\po}$, we select from the look-up table the optimal normalized positive knot positions $\normalizedKnotPosition_k$ and localized difference coefficients $\hat{\approxKsctCoeff}_{kd}$ corresponding to the normalized distance closest to $\adst$, all in floating-point representation.

We then set the knot position quantum counts to
\[\bar{\knotPosition}_{0}=\round{\frac{\knotPosition_{\po}}{\positionalQV}}\text{ and }\;
\bar{\knotPosition}_{k}=\bar{\knotPosition}_{0}+\round{\frac{\pl\normalizedKnotPosition_{k}}{\positionalQV}},\;
\bar{\knotPosition}_{-k}=2\bar{\knotPosition}_{0}-\bar{\knotPosition}_{k}
\]
for \(k=1,\dots,\lastKnotIndex\),
where we use $\round{\cdot}$, a notation inspired by \cite[page 860]{Hastad1989},
to refer to just a usual nearest integer rounding function. In this
context it is irrelevant whether we round $z+\frac{1}{2}$ to $z$
or to $z+1$ for $z\in\mathbb{Z}$.

Then, for $k\ge1$, except for $k=1$ and uneven $d$ if $\numpapprox$
is uneven, the quantized localized difference coefficients are computed as
\[
\bar{\hat{\pwpolACoeff}}_{kd}=\round{\frac{\positionalQV^{d}\pma\scalarField\hat{\approxKsctCoeff}_{kd}}{\coefficientsQV\pd\pl^{d+3}}}
\text{ and }\;
\bar{\hat{\pwpolACoeff}}_{-kd}=\left(-1\right)^{d+1}\bar{\hat{\pwpolACoeff}}_{kd}.
\]

In case of
$\numpapprox$ being even, there is a middle knot with possibly non-zero
coefficients
\[
\bar{\hat{\pwpolACoeff}}_{0d}=-2\sum_{k=-\lastKnotIndex}^{-1}\sum_{j=d}^{\degapprox}\binom{j}{d}\bar{\hat{\pwpolACoeff}}_{kj}\left(\bar{\knotPosition}_{0}-\bar{\knotPosition}_{k}\right)^{j-d}
\]
for uneven $d$.
Otherwise, there is no middle knot, \ie $\bar{\hat{\pwpolACoeff}}_{0d}=0$
for all $d$, but we set
\[
\bar{\hat{\pwpolACoeff}}_{-1d}
=-\!\!\!\!
\sum_{k=-\lastKnotIndex}^{-2}\!\!\!
\bar{\hat{\pwpolACoeff}}_{kd}-\!\!\!
\sum_{j=d+1}^{\degapprox}\!
\binom{\!j}{d\!}\!
\sum_{k=-\lastKnotIndex}^{-1}\!\!\!
\bar{\hat{\pwpolACoeff}}_{kj}\left(\bar{\knotPosition}_{0}-\bar{\knotPosition}_{k}\right)^{j-d}
\]
and $\bar{\hat{\pwpolACoeff}}_{1d}=\left(-1\right)^{d+1}\bar{\hat{\pwpolACoeff}}_{-1d}$ in descending order of uneven $d$.

\subsection{Specifying Quantum Values\label{subsec:specifying-quantum-values}}
Computing optimal quantum values requires the definition of a manageable error measure for minimization. In our attempts to measure the changes to the field approximation originating from quantization, we have developed the data-independent and ray-independent relative quantization error estimate (derivation in the Appendix, \Section \ref{appsec:quantization-error})
\begin{equation*}
    \fc{\QE}{\positionalQV,\coefficientsQV}=\frac{1}{4\knorm}\sqrt{\knormSpecial^{2}\positionalQV^{2}+\sum_{d=0}^{\degapprox}\frac{2\kSubBorder^{2d+3}}{\left(2d+1\right)\left(2d+3\right)}\cdot\frac{\coefficientsQV^{2}}{\positionalQV^{2d}}},
\end{equation*}
where \(\kSubBorder\) is the upper bound of the kernel function's support and we have abbreviated the constants
\begin{align*}
\knorm &= \left(4\pi\int\limits _{t=0}^{\infty}\left[t\fc{\ks}{t}\right]^{2}\diff t\right)^{\frac{1}{2}}\quad\text{and}\\
\knormSpecial &=\left(\intop_{0}^{\kSubBorder}\adst\intop_{\mathbb{R}}\left[\fc{\frac{\diff}{\diff\knotPosition}\ksct{\adst}}{\knotPosition}\right]^{2}\diff\knotPosition\diff\adst\right)^{\frac{1}{2}},
\end{align*}
which only depend on the SPH kernel.

\(\QE\) clearly grows with \(\coefficientsQV\), which is reasonable because the smaller we set \(\coefficientsQV\), the closer the quantized approximations cat get to the optimal ones. However, smaller \(\coefficientsQV\) require larger integer values \(\bar{\pwpolACoeff}_{kd}\) and \(\bar{\hat{\pwpolACoeff}}_{kd}\). Hence, to guard against integer overflow, we propose to set it to the optimal lower bound
\[\coefficientsQV=\frac{\pwpolAmax}{\intmax},\]
where \(\intmax\) is the maximum representable integer for an integer bit-length yet to be chosen, and \(\pwpolAmax\)  an overall upper bound of the values expected to occur in the approximation, which can be generated by a short analysis of the data set to be visualized.

The situation is not as straight-forward for the length quantum \(\positionalQV\). On the one hand, large \(\positionalQV\) result in large quantization errors by distorted knot positions. On the other hand, small \(\positionalQV\) mean large higher-order quantum values \(\coefficientsQV_d\), as we have seen in \Section \ref{subsec:exact-arithmetic-through-quantization}. However, given \(\coefficientsQV\), it is easy to show that \(\fc{\QE}{\positionalQV,\coefficientsQV}\)  is a convex function with respect to \(\positionalQV\), whose minimizer can be found by common root-finding methods (proof in Appendix, \Section \ref{appsec:quantization-error-minimization}).

We have to note, though, while \(\fc{\QE}{\positionalQV,\coefficientsQV}\) estimates the relative quantization error of a normalized particle, the quantization error for a particle from the data set is better represented by \[\fc{\QE}{\frac{1}{\pl}\positionalQV,\frac{\pd\pl^3}{\pma\scalarField}\coefficientsQV},\]
\ie it depends on the particle attributes. 
Thus, for the purpose of determining a suitable global $\positionalQV$, we first specify ``representative'' attributes \(\representative{\pma}\), \(\representative{\pd}\), \(\representative{\pl}\), and \(\representative{\scalarField}\) of the data set to be visualized. These could be, for example, average or mean values or the attributes chosen from any particle in a region of interest. Afterwards, abbreviating $\representative{\pcmax}=\frac{\vphantom l\representative{\pma}\representative{\scalarField}}{\smash[b]{\representative{\pd}\representative{\pl}^{3}}}$, we set $\positionalQV$ to minimize  
$\fc{\QE}{\frac{\positionalQV}{\representative{\pl}},\frac{\coefficientsQV}{\representative{\pcmax}}}$.

\section{\uppercase{Choosing Approximation Dimensions}\label{sec:choosing-approximation-dimensions}}
Having set forth our quantized higher-order approximation field concept, the question remains how to choose its most fundamental parameters: the approximation order \(\degapprox\), the number of non-trivial polynomial pieces per particle \(\numpapprox\), and the bit length defining the maximum representable integer \(\intmax\). All three parameters can significantly impact both approximation accuracy and performance. While the quality of any configuration \(\left(\degapprox,\numpapprox,\intmax\right)\) ultimately requires thorough testing to be readily evaluated, we do want to provide an overview in theory here.

\subsection{Performance Implications}
The method's asymptotic complexity with respect to \(\numpapprox\) and \(\degapprox\) is easily seen: \(\numpapprox\) has linear impact on the number of knots per ray and therefore a linearithmic one on the overall process time due to the search sweep. The overall time effect of \(\degapprox\) is quadratic since both the number of coefficients to be updated and the number of operations for each coefficient update according to (\ref{eq:accumulation-rule}) grow linearly with \(\degapprox\).

The performance implications of the integer bit length require special attention. In spite of the focus on floating-point performance, modern GPUs intrinsically support calculations on integer types of 16-bit and 32-bit lengths. For 32-bit integers, the throughput of additions is comparable to 32-bit floating-point operations while multiplications are commonly processed about five times slower. 64-bit integer operations are formally supported in some GPU programming contexts (Vulkan API, OpenCL, CUDA, extension of GLSL) but seem to be always emulated in software based on 32-bit operations.

Such an emulation can easily be constructed for integer types of any size, which means that in theory we can go with arbitrarily small quantum values, albeit at a high price performance-wise. 
To get an impression about the performance implications of using such long integers, we have analysed such algorithms. Compared to 32-bit integers, our results show a cost increase by a factor of roughly 9 for bit-length 64, 24 for 96 bit, and 45 for 128 bit, although we expect 64-bit integers to be especially well-optimized in the implementations by GPU vendors.

To be able to decide whether these cost factors are worthwhile and, consequently, choose adequate quantum values, we have to form an idea of the quality implications of smaller or larger integer sizes. In other words, we have to relate the quantization cost to the quantization error.

\subsection{Combined Accuracy Measure}
In order to estimate the overall accuracy implications of \(\degapprox\), \(\numpapprox\), and \(\intmax\), we combine the error estimate for piecewise polynomial approximation and the one for quantization.

In \Section \ref{subsec:optimization-problem-definition}, we have already defined the error \(\fc{\approxError}{\approxKsct}\) for a non-quantized approximation along a ray with normalized distance \(\adst\). To become independent from \(\adst\)  and the screen resolution, we integrate the optimal per-ray error over all viewing rays in one direction. We then divide it by the \(L^2\) norm \(\knorm\) of a normalized particle's contribution to arrive at the per-particle relative error for higher-order approximation
\[
\overallApproxError=\frac{1}{\knorm}\left(2\pi\intop_{0}^{\kSubBorder}\adst\fc{\approxError^2}{\bestApproxKsct}\diff\adst\right)^\frac{1}{2},
\]
where we have used \(\bestApproxKsct\) to refer to the optimal approximation at distance \(\adst\) with optimal knot positions. As we are given these approximations only implicitly through a minimization process, we compute \(\overallApproxError\) only approximately as a Riemann sum. It constitutes a precise relative approximation error measure for any particle, not just for the normalized ones.

By contrast, the quantization error is difficult to quantify exactly, due to the statistical and rather complex distortion caused by quantization. We thus content ourselves with the rather rough estimate \(\fc{\QE}{\frac{\positionalQV}{\representative{\pl}},\frac{\coefficientsQV}{\representative{\pcmax}}}\) defined in \Section \ref{subsec:specifying-quantum-values}. While this measure is already in the form of a per-particle relative error derived from an all-rays integration similar to the one for \(\overallApproxError\) above, it has the disadvantage of apparently depending on \(\positionalQV\), \(\coefficientsQV\), and the particle data. Fortunately, a closer look reveals that the only property of the data set that \(\QE\) really depends on is
\[\frac{\pwpolAmax}{\representative{\pcmax}},\]
\ie the ratio between a ``representative'' particle's factor to the SPH kernel and an upper bound of the target field value. It is a measure for the variance of the target scalar field. Fixing this value at, say, \(10^5\) to be robust against integer overflow for at least some level of data variance, we can compute \(\frac{\coefficientsQV}{\representative{\pcmax}}\) and thus the optimal ratio \(\frac{\positionalQV}{\representative{\pl}}\) providing the error \(\QE\).

Committed on a quantization error estimate \(\QE\), we can combine it to \(\overallApproxError\) above to form an overall approximation accuracy measure. Simply summing the two errors would introduce an overestimation bias as it would model all distortions acting in the same direction. Instead, we treat the two sources of error as if they were perpendicular and take the \(L^2\) norm of their sum, arriving at the overall approximation error
\[\sqrt{\overallApproxError^2+\QE^2}\]
for evaluating configurations \(\left(\degapprox,\numpapprox,\intmax\right)\).

\begin{figure}[H]
\centering
\pgfplotsset{every axis/.append style={
xmin=0.45,
xmax=6.55,
ymode=log,
log origin=infty,
ymin=\boxplotymin,
ymax=0.13,
height=8cm,
ymajorgrids,
tick style={black!40},
tick label style={inner sep=1pt},
axis x line=none,
axis y line*=left,
mark=-,
mark size=0.2cm,
bar width=0.4cm,
width=4.9cm,
}}
\pgfmathsetmacro{\boxplotymin}{0.0000031}
\begin{tikzpicture}[
	p1/.style={ultra thick,mygreen,line cap=round,inner sep=2pt
},
	p2/.style={ultra thick,myred,line cap=round,inner sep=2pt
},
	p3/.style={ultra thick,myblue,line cap=round,inner sep=2pt
},
	p4/.style={ultra thick,myorange,line cap=round,inner sep=2pt
},
	q/.style={fill=black!25,draw=none},
	qbg/.style={fill=black!5,draw=none},
]

\begin{scope}[]
\node at (1.7,6.8) {$\overallApproxError$};
\begin{axis}[
axis y line*=right,
yticklabels=\empty
]
\coordinate (origin4) at (0.5,\boxplotymin);
\addplot[ybar,qbg] coordinates{ (1, 1) (2, 1) (3, 1) (4, 1) (5, 1) (6, 1) };
\addplot[p1,only marks] coordinates{
(2, 0.106316306613374) (4, 0.0533987770065947) (6, 0.00853283332898273)
};
\addplot[p2,only marks] coordinates{
(1, 0.0579492894294634) (2, 0.0577694758801597) (3, 0.0270944310632377) (4, 0.00454428903348760) (5, 0.00322446021459762) (6, 0.00133487883301723)
};
\addplot[p3,only marks] coordinates{
(1, 0.0476217249112189) (2, 0.0110676106608898) (3, 0.0107138845645221) (4, 0.00101729884209219) (5, 0.000592716195927300) (6, 0.000170378565776039)
};
\addplot[p4,only marks] coordinates{
(1, 0.0455525784862875) (2, 0.00827470955036559) (3, 0.000925355941451358) (4, 0.000166750209347242) (5, 0.0000324814641680290) (6, 6.92753461586753e-6)
};
\path \foreach \x in {1,...,6} { (\x,\boxplotymin) node[anchor=south] {$\x$} };
\end{axis}
\node[anchor=north west] (label4) at (origin4) {$\degapprox$};
\draw[->] (label4.east) --+(0.5,0);
\end{scope}

\begin{scope}[xshift=4.1cm]
\node at (1.7,6.8) {$\sqrt{\overallApproxError^2+\QE^2}$};
\begin{axis}
\coordinate (origin2) at (0.5,\boxplotymin);
\addplot[ybar,qbg] coordinates{ (1, 1) (2, 1) (3, 1) (4, 1) (5, 1) (6, 1) };
\addplot[ybar,q] coordinates{

(2, 6.30864792448258e-6) (3, 0.0000855541558867942) (4, 0.000410901829274932) (5, 0.00117290391232476) (6, 0.00248526771799010)
};
\addplot[p1,only marks] coordinates{
(2, 0.106316306800547) (4, 0.0534003579212099) (6, 0.00888739558308744)  
};
\addplot[p2,only marks] coordinates{
(1, 0.0579492894294740) (2, 0.0577694762246239) (3, 0.0270945661370344) (4, 0.00456282841373387) (5, 0.00343115829758256) (6, 0.00282107375460500)
};
\addplot[p3,only marks] coordinates{
(1, 0.0476217249112318) (2, 0.0110676124588856) (3, 0.0107142261491670) (4, 0.00109714960120468) (5, 0.00131415983672507)
};
\addplot[p4,only marks] coordinates{
(1, 0.0455525784863010) (2, 0.00827471195522540) (3, 0.000929302497558697) (4, 0.000443447793566317) (5, 0.00117335358398960)
};
\coordinate (c36) at (6, 0.00249110105892146);
\coordinate (c46) at (6, 0.00248527737301485);
\path \foreach \x in {1,...,6} { (\x,\boxplotymin) node[anchor=south] {$\x$} };
\end{axis}
\node[anchor=north west] (label2) at (origin2) {$\degapprox$};
\draw[->] (label2.east) --+(0.5,0);
\draw[p3] (c36)+(-0.025,0)--+(-0.2,0);
\draw[p4] (c46)+(0.025,0)--+(0.2,0);
\end{scope}

\begin{scope}[yshift=-1.2cm,xshift=0.8cm]
\draw[p1] (0,0.2) --+(1,0);
\draw[p2] (1.6,0.2) --+(1,0);
\draw[p3] (3.2,0.2) --+(1,0);
\draw[p4] (4.8,0.2) --+(1,0);
\node[anchor=south,p1] at (0.5,0.3) {$\numpapprox=1$};
\node[anchor=south,p2] at (2.1,0.3) {$\numpapprox=2$};
\node[anchor=south,p3] at (3.7,0.3) {$\numpapprox=3$};
\node[anchor=south,p4] at (5.3,0.3) {$\numpapprox=4$};
\end{scope}

\end{tikzpicture}

\caption[combined approximation error for cubic B-spline kernel]{\label{fig:combined-approximation-error}
Plot of error values in the example case of the cubic B-Spline SPH kernel (\ref{eq:cubic-kernel}), for $\degapprox \leq 6$ and $\numpapprox\leq 4$.
The horizontal colored marks show the polynomial approximation error component $\overallApproxError$ on the left and the combined error $\sqrt{\overallApproxError^2+\QE^2}$ on the right. The grey bars on the right-hand side depict the quantization error component $\QE$, which has been computed assuming an integer bit length of 64 and a data variance factor  $\nicefrac{\pwpolAmax}{\representative{\pcmax}}=10^5$.
}
\end{figure}
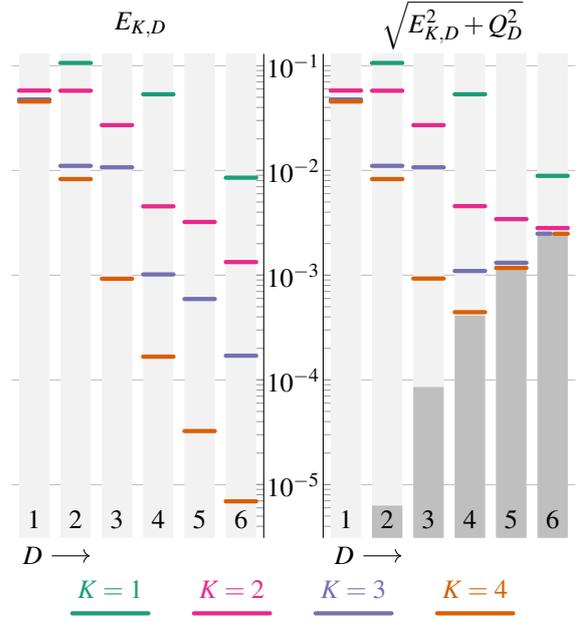

The overall error clearly falls with growing \(\numpapprox\)  and \(\intmax\) as these two parameters only effect one of the two components.
However, the effect of \(\degapprox\) is more interesting because higher \(\degapprox\) cause \(\overallApproxError\) to fall but \(\QE\) to grow. Thus, for any fixed \(\numpapprox\) and \(\intmax\), there is one error-minimizing \(\degapprox\), such that raising the approximation order further will not be worthwhile as it will reduce the accuracy at even higher cost. \Figure \ref{fig:combined-approximation-error} shows a plot of \(\overallApproxError\) and \(\QE\), as well as the combined error, for the cubic B-spline kernel (\ref{eq:cubic-kernel}). It covers values for approximation order \(\degapprox\) up to 6 and per-particle non-trivial pieces count \(\numpapprox\) up to 4, assuming an integer bit length of 64 and a data variance ratio of $\nicefrac{\pwpolAmax}{\representative{\pcmax}}=10^5$. One can see that for, \eg  \(\numpapprox=3\) and \(\numpapprox=4\), raising \(\degapprox\) above 4 is clearly not worthwhile.

While it is hard to select one universal configuration from the analysis presented here, a visualization tool using our method can implement a single one or several, and even more than one SPH kernel function. For each configuration, \(\overallApproxError\) can be computed at compile time. Also \(\QE\), being a one-dimensional function of $\nicefrac{\pwpolAmax}{\representative{\pcmax}}$ for each \(\degapprox\) and \(\intmax\) after all, can quickly be obtained from a look-up table, facilitating an efficient computation of the combined error estimate when loading a data set. 
Thus, although choosing an ideal configuration depends on user preferences for balancing quality and speed, a visualization tool may restrict a set of implemented configurations to a preselection of worthwhile configurations in view of the target data set for the user to choose from.

\section{\uppercase{Conclusion}\label{sec:conclusion}}

Seeking a new level of quantitative accuracy in scientific visualization, we have presented a novel approach to approximating SPH scalar fields on viewing rays during volume ray casting. It features a locally adaptive spatial resolution and efficient summation scheme.
We have shown how to efficiently compute the best possible higher-order approximations of particle contributions of any given order and resolution, and provided a thorough theoretic analysis of the approximation errors involved. Conveying these error estimates to the user, could meet a field expert's need for quantitatively assessing the errors involved in the visualization process.

While we have confined our explanations on representing field values, the procedure for gradients or other field features is analogous, as long as these are additively generated from particle contributions. Also, despite our focus on SPH data, our concepts may very well be applicable to direct volume renderings of scattered data.

Clearly, our findings have yet to prove their competitiveness in practise. However, we are confident that they will help in advancing scientific visualization of scattered data in terms of quantitative accuracy.

\bibliographystyle{apalike}
{\small
\bibliography{references}}

\begin{thebibliography}{}

\bibitem[Drebin et~al., 1988]{Drebin1988}
Drebin, R.~A., Carpenter, L., and Hanrahan, P. (1988).
\newblock Volume rendering.
\newblock In {\em ACM Siggraph Computer Graphics}, volume 22.4, pages 65--74. ACM.

\bibitem[Fraedrich et~al., 2010]{Fraedrich2010}
Fraedrich, R., Auer, S., and Westermann, R. (2010).
\newblock Efficient high-quality volume rendering of {SPH} data.
\newblock {\em IEEE Transactions on Visualization and Computer Graphics}, 16(6):1533--1540.

\bibitem[Gingold and Monaghan, 1977]{Gingold1977}
Gingold, R.~A. and Monaghan, J.~J. (1977).
\newblock Smoothed particle hydrodynamics: theory and application to non-spherical stars.
\newblock {\em Monthly notices of the royal astronomical society}, 181(3):375--389.

\bibitem[Hastad et~al., 1989]{Hastad1989}
Hastad, J., Just, B., Lagarias, J.~C., and Schnorr, C.-P. (1989).
\newblock Polynomial time algorithms for finding integer relations among real numbers.
\newblock {\em SIAM Journal on Computing}, 18(5):859--881.

\bibitem[Hochstetter et~al., 2016]{hochstetter2016adaptive}
Hochstetter, H., Orthmann, J., and Kolb, A. (2016).
\newblock Adaptive sampling for on-the-fly ray casting of particle-based fluids.
\newblock In {\em Proceedings of High Performance Graphics}, pages 129--138. The Eurographics Association.

\bibitem[K{\"a}hler et~al., 2007]{Kahler2007}
K{\"a}hler, R., Abel, T., and Hege, H.-C. (2007).
\newblock Simultaneous gpu-assisted raycasting of unstructured point sets and volumetric grid data.
\newblock In {\em Proceedings of the Sixth Eurographics/Ieee VGTC conference on Volume Graphics}, pages 49--56. Eurographics Association.

\bibitem[Knoll et~al., 2014]{Knoll2014}
Knoll, A., Wald, I., Navratil, P., Bowen, A., Reda, K., Papka, M.~E., and Gaither, K. (2014).
\newblock Rbf volume ray casting on multicore and manycore cpus.
\newblock In {\em Computer Graphics Forum}, volume 33.3, pages 71--80. Wiley Online Library.

\bibitem[Lucy, 1977]{Lucy1977}
Lucy, L.~B. (1977).
\newblock A numerical approach to the testing of the fission hypothesis.
\newblock {\em The astronomical journal}, 82:1013--1024.

\bibitem[Max, 1995]{Max1995}
Max, N. (1995).
\newblock Optical models for direct volume rendering.
\newblock {\em IEEE Transactions on Visualization and Computer Graphics}, 1(2):99--108.

\bibitem[Oliveira and Paiva, 2022]{oliveira2022narrow}
Oliveira, F. and Paiva, A. (2022).
\newblock Narrow-band screen-space fluid rendering.
\newblock In {\em Computer Graphics Forum}, volume 41.6, pages 82--93. Wiley Online Library.

\bibitem[Rosswog, 2009]{Rosswog2009}
Rosswog, S. (2009).
\newblock Astrophysical smooth particle hydrodynamics.
\newblock {\em New Astronomy Reviews}, 53(4-6):78--104.

\end{thebibliography}

\vfill
\section*{\uppercase{Appendix}}
\renewcommand{\thesubsection}{\arabic{subsection}}
\begin{figure*}[!b]
  \centering
  \begin{tikzpicture}[thick]

\tikzmath{
\b1=1;
\b2=1.7;
\b3=3;
\plotend=3.5;
function basef(\x,\k,\d){
\thisbmo=0;
\thisb=\b1;
if \k==2 then {\thisbmo=\b1; \thisb=\b2; };
if \k==3 then {\thisbmo=\b2; \thisb=\b3; };
\ax=abs(\x);
if \ax < \thisbmo then { return 1; } else {
if \ax > \thisb then { return 0; } else {
return 1 - ((\ax - \thisbmo) / (\thisb - \thisbmo))^\d; }; };
};
}
\foreach \k in {1,...,3} \foreach \d in {1,...,4}{
\begin{axis}[
 tiny,
 xmin=-\plotend,
 xmax=\plotend,
 ymin=-0.3,
 ymax=1.3,
 width=4cm,
 height=0.8cm,
 scale only axis,
 xshift=\k * 4.65cm - 4.65cm,
 yshift=1.45cm -\d * 1.45 cm,
 ytick={0,1},
 xtick={-\b3,-\b2,-\b1,0,\b1,\b2,\b3},
 xticklabels={$-\normalizedKnotPosition_{3}\;\,$,$-\normalizedKnotPosition_{2}\quad\;$,$-\normalizedKnotPosition_{1}$,$0$,$\;\normalizedKnotPosition_{1}$,$\;\,\normalizedKnotPosition_{2}$,$\;\,\normalizedKnotPosition_{3}$},
 ]
\addplot[thick, myred,domain=-\plotend:\plotend,samples=141]{basef(x,\k,\d)};
\end{axis}
}
\fill[white,opacity=0.6](-0.3,-0.4) rectangle +(4.5,1.4) ++(0,-2.9) rectangle +(4.5,1.4);
\foreach \k in {1,...,3} \node[inner sep=0] at (\k * 4.65 - 2.65,1.2) {$k=\k$};
\foreach \d in {1,...,4} \node[inner sep=0,anchor=west] at (-1.5,1.85 - \d * 1.45) {$d=\d$};

\end{tikzpicture}
  \caption{Plots of the basis functions $\nonorthogbasiselement kd$ for $k=1,\dots,3$, $d=1,\dots,4$, and an arbitrary setting of positive knot positions $\normalizedKnotPosition_{1}$, $\normalizedKnotPosition_{2}$, and $\normalizedKnotPosition_{3}$. These 12 functions are the elements of the basis $\nonorthogbasis$ for $\numpapprox=6$ and $\degapprox=4$. For $\numpapprox=5$, there are still three positive knot positions but the elements $\nonorthogbasiselement 11$ and $\nonorthogbasiselement 13$ drawn in paler colors are excluded because they feature a polynomial change at $\knotPosition=0$.}
  \label{fig:basis-function-plots}
 \end{figure*}

\subsection{Basis of the Set of Eligible Approximation Candidates for Fixed Knot Positions\label{appsec:basis}}

Let
\begin{itemize}
    \item $\numpapprox$ and $\degapprox$ be positive integers,
    \item $0=\normalizedKnotPosition_{0}<\normalizedKnotPosition_{1}<\dots<\normalizedKnotPosition_{\lastKnotIndex}$ a growing real sequence,
    \item $\approxKsctSet$ the set of even continuous piecewise polynomial functions with compact support, maximal degree \(\degapprox\), maximal number of non-trivial pieces \(\numpapprox\), and positive knot positions \(\normalizedKnotPosition_{1},\dots,\normalizedKnotPosition_{\lastKnotIndex}\),
    \item  $\orthogbasisIndexpairSet$ the set of pairs $\left(k,d\right)$ of positive integers $k\le\lastKnotIndex$
and $d\le\degapprox$ but excluding elements $\left(1,d\right)$ for
uneven $d$ if $\numpapprox$ is uneven.
\end{itemize}
Then the set \[\fc{\nonorthogbasis}{\normalizedKnotPosition_{1},\dots,\normalizedKnotPosition_{\lastKnotIndex}}=\left\{ \vphantom{\big\{}\smash{\nonorthogbasiselement kd}:\left(k,d\right)\in\orthogbasisIndexpairSet\right\} \]
of functions
\[
\fc{\nonorthogbasiselement kd}{\knotPosition}=\begin{cases}
1 & \text{if }\left|\knotPosition\right|\le\normalizedKnotPosition_{k-1}\\
1-\left(\frac{\left|\knotPosition\right|-\normalizedKnotPosition_{k-1}}{\normalizedKnotPosition_{k}-\normalizedKnotPosition_{k-1}}\right)^{d} & \text{if }\normalizedKnotPosition_{k-1}\leq\left|\knotPosition\right|\leq\normalizedKnotPosition_{k}\\
0 & \text{if }\left|\knotPosition\right|\ge\normalizedKnotPosition_{k}
\end{cases}
\]
is a basis of $\fc{\approxKsctSet}{\normalizedKnotPosition_{1},\dots,\normalizedKnotPosition_{\lastKnotIndex}}$.

To illustrate the structure of this basis, \Figure\ref{fig:basis-function-plots}
shows plots of the graphs of the elements of $\fc{\nonorthogbasis}{\normalizedKnotPosition_{1},\normalizedKnotPosition_{2},\normalizedKnotPosition_{3}}$
for $\lastKnotIndex=3$ and $\degapprox=4$.
\begin{proof}
By construction, all $\nonorthogbasiselement kd$ are clearly even
continuous polynomial functions with compact support, maximal degree
$\degapprox$ and positive knot positions $\normalizedKnotPosition_{1},\dots,\normalizedKnotPosition_{\lastKnotIndex}$.
They comprise not more than $\numpapprox$ non-trivial pieces, including
for uneven $\numpapprox$, because then there is no change of the polynomial
at $\knotPosition=\normalizedKnotPosition_{0}$ since, for even $d$
and because $\normalizedKnotPosition_{0}=0$,
\[
1-\left(\frac{\left|\knotPosition\right|-\normalizedKnotPosition_{0}}{\normalizedKnotPosition_{1}-\normalizedKnotPosition_{0}}\right)^{d}=1-\frac{\left(-1\right)^{d}\knotPosition^{d}}{\normalizedKnotPosition_{1}^{d}}=1-\frac{\knotPosition^{d}}{\normalizedKnotPosition_{1}^{d}}.
\]
 It therefore suffices to show that $\fc{\nonorthogbasis}{\normalizedKnotPosition_{1},\dots,\normalizedKnotPosition_{\lastKnotIndex}}$
is linearly independent and generates the entire vector space.

For this proof, we use
the indexed bracket notation 
\[
\coeffofk dk{\pwpolA}
\]
for a piecewise polynomial $\pwpolA$ with positive knot positions
$\normalizedKnotPosition_{1},\dots,\normalizedKnotPosition_{\lastKnotIndex}$,
$d\ge0$, and $k\le\lastKnotIndex$, to denote the coefficient of
order $d$ of the polynomial corresponding to the subdomain $\left[\normalizedKnotPosition_{k-1},\normalizedKnotPosition_{k}\right]$.

We start off with two simple observations about some of the coefficients
of the basis functions $\nonorthogbasiselement kd$, firstly,
\begin{align}
\coeffofk dk{\nonorthogbasiselement kd}& =\coeffofk dk{1-\left(\frac{\left|\knotPosition\right|-\normalizedKnotPosition_{k-1}}{\normalizedKnotPosition_{k}-\normalizedKnotPosition_{k-1}}\right)^{d}}\nonumber \\
&=-\left(\normalizedKnotPosition_{k}-\normalizedKnotPosition_{k-1}\right)^{-d}\label{eq:non-orthog-proof-first-observation}
\end{align}
 for all $\left(k,d\right)\in\orthogbasisIndexpairSet$ and, secondly,
\begin{equation}
k^{*}\neq k\;\vee\;d^{*}>d\quad\Longrightarrow\quad\coeffofk{d^{*}}{k^{*}}{\nonorthogbasiselement kd}=0.\label{eq:non-orthog-proof-second-observation}
\end{equation}

Now, to show linear independence, let $a_{kd}\in\mathbb{R}$ for all
$\left(k,d\right)\in\orthogbasisIndexpairSet$ and $\sum_{\left(k,d\right)\in\orthogbasisIndexpairSet}a_{kd}\nonorthogbasiselement kd\equiv\mathbf{0}$.
For a proof by contradiction, suppose there is a non-empty subset
$\orthogbasisIndexpairSet'\subseteq\orthogbasisIndexpairSet$ such
that $a_{kd}\neq0$ for all $\left(k,d\right)\in\orthogbasisIndexpairSet'$.
Since $\orthogbasisIndexpairSet'$ is finite, it contains one maximal
element $\left(k^{*},d^{*}\right)$ with respect to lexicographical
order, \ie such that all $\left(k,d\right)\in\orthogbasisIndexpairSet'\setminus\left\{ \left(k^{*},d^{*}\right)\right\} $
satisfy $k<k^{*}\;\vee\;\left(k=k^{*}\;\wedge\;d<d^{*}\right)$. The
above observations (\ref{eq:non-orthog-proof-first-observation})
and (\ref{eq:non-orthog-proof-second-observation}) then imply 
\begin{align*}
\coeffofk{d^{*}}{k^{*}}{\sum_{\left(k,d\right)\in\orthogbasisIndexpairSet}a_{kd}\nonorthogbasiselement kd}
={}& a_{k^{*}d^{*}}\left(\coeffofk{d^{*}}{k^{*}}{\nonorthogbasiselement{k^{*}}{d^{*}}}\right) \\
&+\sum_{\substack{\left(k,d\right)\in \\ \orthogbasisIndexpairSet'\setminus\left\{ \left(d^{*},k^{*}\right)\right\} }}
a_{kd}\left(\coeffofk{d^{*}}{k^{*}}{\nonorthogbasiselement kd}\right)\\
={}& -a_{k^{*}d^{*}}\left(\normalizedKnotPosition_{k^{*}}-\normalizedKnotPosition_{k^{*}-1}\right)^{-d^{*}}\\
\neq{}& 0,
\end{align*}
 which proves that the linear combination cannot be zero.

To prove the generator property, let $\approxKsct\in\fc{\approxKsctSet}{\normalizedKnotPosition_{1},\dots,\normalizedKnotPosition_{\lastKnotIndex}}$.
We now show how to generate multipliers $a_{kd}$ such that the corresponding
linear combination 
\[
\pwpolA=\sum_{\left(k,d\right)\in\orthogbasisIndexpairSet}^{\degapprox}a_{kd}\nonorthogbasiselement kd
\]
 is equal to $\approxKsct$.

For all $k^{*}=2,\dots,\lastKnotIndex$, we recursively set the multipliers
$a_{k^{*}d^{*}}$ for decreasing $d^{*}=\degapprox,\dots,1$ by the
rule 
\begin{equation*}
\begin{split}
%\begin{align*}
a_{k^{*}d^{*}}= & \left(\normalizedKnotPosition_{k^{*}}-\normalizedKnotPosition_{k^{*}-1}\right)^{d^{*}} \\
& \cdot
\left[\left(\coeffofk{d^{*}}{k^{*}}{\sum_{d=d^{*}+1}^{\smash{\degapprox}}a_{k^{*}d}\nonorthogbasiselement{k^{*}}d}\right)-\left(\coeffofk{d^{*}}{k^{*}}{\approxKsct}\right)\right],
%\end{align*}
\end{split}
\end{equation*}
 which is well-defined because the $a_{k^{*}d}$ for $d^{*}+1\le d\le\degapprox$
have already been set before. By (\ref{eq:non-orthog-proof-first-observation})
and (\ref{eq:non-orthog-proof-second-observation}), this then guarantees
\begin{align*}
\coeffofk{d^{*}}{k^{*}}{\pwpolA} ={}& \left(\coeffofk{d^{*}}{k^{*}}{\sum_{d=d^{*}}^{\smash{\degapprox}}a_{k^{*}d}\nonorthogbasiselement{k^{*}}d}\right) \\
& +\left(\coeffofk{d^{*}}{k^{*}}{\sum_{k\smash{\neq}k^{*}\wedge d<d^{*}}a_{kd}\nonorthogbasiselement kd}\right)\\
={} & -a_{k^{*}d^{*}}\left(\normalizedKnotPosition_{k^{*}}-\normalizedKnotPosition_{k^{*}-1}\right)^{-d^{*}}\\
& +\left(\coeffofk{d^{*}}{k^{*}}{\sum_{d=d^{*}+1}^{\smash{\degapprox}}a_{k^{*}d}\nonorthogbasiselement{k^{*}}d}\right)\vphantom{\sum_{d}^{\degapprox}}\\
={} & \coeffofk{d^{*}}{k^{*}}{\approxKsct}
\end{align*}
 for any positive $d^{*}$ and $k^{*}\ge2$.

For $k^{*}=1$ and decreasing $d^{*}$, excluding uneven $d^{*}$
in the case of uneven $\numpapprox$, we adhere to the same rule,
which now amounts to 
\[
a_{1d^{*}}=-\normalizedKnotPosition_{1}^{d^{*}}\cdot\coeffofk{d^{*}}1{\approxKsct}
\]
 because $\nonorthogbasiselement 1{d^{*}}=1-\left(\frac{\knotPosition}{\normalizedKnotPosition_{1}}\right)^{d^{*}}$
for $\knotPosition\in\left[0,\normalizedKnotPosition_{1}\right]$.
This guarantees $\coeffofk{d^{*}}1{\pwpolA}=\coeffofk{d^{*}}1{\approxKsct}$
at least for even $\numpapprox$ or even $d^{*}$. For uneven $\numpapprox$
and $d^{*}$, the equality holds, too, because $\approxKsct$ may
not change the polynomial at $\normalizedKnotPosition_{0}=0$ and
$\restr{\approxKsct}{\left[-\normalizedKnotPosition_{1},\normalizedKnotPosition_{1}\right]}$
must be even, requiring all of its uneven coefficients to vanish,
which is also the case for the elements of $\fc{\nonorthogbasis}{\normalizedKnotPosition_{1},\dots,\normalizedKnotPosition_{\lastKnotIndex}}$
by construction.

Therefore, as the coefficients of $\pwpolA$ and $\approxKsct$ in
all subdomains intersecting the positive real axis match for positive
order, their difference $\approxKsct-\pwpolA$ must be piecewise constant
there. Since $\approxKsct-\pwpolA$ is also even and continuous and
has compact support, due to $\approxKsct$ and A having the same properties,
it must be the zero function.
\end{proof}
\subsection{Proof of Optimality of $\bestApproxKsct$\label{appsec:optimality}}
Let \(L\) be a Hilbert space with scalar product 
\(\left\langle \squareIntegrableA,\squareIntegrableB\right\rangle\) and induced norm \(\left\Vert \squareIntegrableA\right\Vert=\sqrt{\left\langle \squareIntegrableA,\squareIntegrableA\right\rangle }\) for any $ \squareIntegrableA, \squareIntegrableB \in L$.
For \(\squareIntegrableA,\squareIntegrableB\in L\), \(\squareIntegrableB\not\equiv\mathbf{0}\), we denote by \[\proj{\squareIntegrableB}{\squareIntegrableA}=\frac{\left\langle \squareIntegrableA,\squareIntegrableB\right\rangle }{\left\langle \squareIntegrableB,\squareIntegrableB\right\rangle }\squareIntegrableB\]
the orthogonal projection of \(\squareIntegrableA\) on \(\squareIntegrableB\).

Further, let \(\approxKsctSet\) be a vector subspace of \(L\) of finite dimension with orthogonal basis \(\orthogbasis\).

Then for any $\ksct{\adst}\in L$, its orthogonal projection on $\approxKsctSet$,
\begin{equation}
\label{eq:optimal-approx-for-given-knots-appendix}
\bestApproxKsct=\sum_{\pwpolA\in\orthogbasis}\proj{\pwpolA}{\ksct{\adst}}
\end{equation}
is the unique element in $\approxKsctSet$ with minimal norm $\left\Vert \bestApproxKsct-\ksct{\adst}\right\Vert$.

\begin{proof}
By construction, $\bestApproxKsct$ is a linear combination of the elements of basis $\orthogbasis$, hence $\bestApproxKsct\in\approxKsctSet$.

For any other element $\approxKsct\in\approxKsctSet$,
\[
\left\langle \ksct{\adst}-\bestApproxKsct,\approxKsct-\bestApproxKsct\right\rangle =0,
\]
due to (\ref{eq:optimal-approx-for-given-knots-appendix}) and the elements of $\orthogbasis$ being mutually orthogonal. Hence, in case of $\approxKsct\neq\bestApproxKsct$,
\begin{align*}
\left\Vert \approxKsct-\ksct{\adst}\right\Vert^{2}={}& \left\langle \approxKsct-\ksct{\adst},\approxKsct-\ksct{\adst}\right\rangle \\
={}& \left\langle \approxKsct-\bestApproxKsct+\bestApproxKsct-\ksct{\adst},\approxKsct-\bestApproxKsct+\bestApproxKsct-\ksct{\adst}\right\rangle \\
={}& \left\langle \approxKsct-\bestApproxKsct,\approxKsct-\bestApproxKsct\right\rangle +2\left\langle \approxKsct-\bestApproxKsct,\bestApproxKsct-\ksct{\adst}\right\rangle \\
&+\left\langle \bestApproxKsct-\ksct{\adst},\bestApproxKsct-\ksct{\adst}\right\rangle \\
={}& \left\Vert \approxKsct-\bestApproxKsct\right\Vert^{2}+2\cdot0+\left\Vert \bestApproxKsct-\ksct{\adst}\right\Vert^{2}\\
>{}& \left\Vert \bestApproxKsct-\ksct{\adst}\right\Vert^{2}.
\end{align*}
\end{proof}

\subsection{Derivation of Computation Rules for Quantized Knots\label{appsec:quantized-knots}}

Given a particle with attributes $\pma$, $\pd$, $\pl$,  and $\scalarField$,  positioned at distance $\adst\pl$ from the ray with closest point parameter $\knotPosition_{\po}$, we select from the look-up table the optimal normalized positive knot positions $\normalizedKnotPosition_k$ and localized difference coefficients $\hat{\approxKsctCoeff}_{kd}$ corresponding to the normalized distance closest to $\adst$, all in floating-point representation.
They define an optimal approximation $\bestApproxKsct$ of $\ksct\adst$, such that the given particle's contribution is approximated by
the (unquantized) piecewise polynomial function
\[
\fc{\approxFromLUT}{\knotPosition}=\frac{\pma\sf}{\pd\pl^{3}}\fc{\approxKsct_{\normadst}}{\frac{\knotPosition-\knotPosition_{\po}}{\pl}}.
\]

For given quantum values $\positionalQV$ and $\coefficientsQV$,
our goal is now to compute quantized knots
\[
\left(\bar{\knotPosition}_{k},\bar{\hat{\approxCandCoeff}}_{k0},\dots,\bar{\hat{\approxCandCoeff}}_{k\degapprox}\right),\bar{\hat{\approxCandCoeff}}_{k0}=0,\text{ for }k=-\lastKnotIndex,\dots,\lastKnotIndex
\]
 encoding the continuous contribution approximation
\begin{equation}
\fc{\approxCandidate}{\knotPosition}=\begin{cases}
\fc{\approxCandPiece{-\lastKnotIndex-1}}{\knotPosition} & \text{if }\knotPosition\le\bar{\knotPosition}_{-\lastKnotIndex}\positionalQV\\[3pt]
\fc{\approxCandPiece k}{\knotPosition} & \begin{aligned}
  & \text{if }\bar{\knotPosition}_{k}\positionalQV\le\knotPosition\le\bar{\knotPosition}_{k+1}\positionalQV \text{ for some } \\[-3pt] & k\in\left\{ -\lastKnotIndex,\dots,\lastKnotIndex-1\right\} \end{aligned}\\[3pt]
\fc{\approxCandPiece{\lastKnotIndex}}{\knotPosition} & \text{if }\knotPosition\ge\bar{\knotPosition}_{\lastKnotIndex}\positionalQV
\end{cases}\label{eq:contribution-spline-formula}
\end{equation}
 through 
\[
\approxCandPiece{-\lastKnotIndex-1}\equiv\v 0
\]
and 
\begin{equation}
\fc{\approxCandPiece k}{\knotPosition}-\fc{\approxCandPiece{k-1}}{\knotPosition}=\coefficientsQV\sum_{d=0}^{\degapprox}\bar{\hat{\approxCandCoeff}}_{kd}\left(\knotPosition-\bar{\knotPosition}_{k}\positionalQV\right)^{d}
\label{eq:quantized-localized-diff-coord-definition}
\end{equation}
for $k=-\lastKnotIndex,\dots,\lastKnotIndex$
such that the quantized localized coefficients $\bar{\approxCandCoeff}_{kd}$
of the individual polynomial functions 
\[
\fc{\approxCandPiece k}{\knotPosition}=\coefficientsQV\sum_{d=0}^{\degapprox}\bar{\approxCandCoeff}_{kd}\left(\knotPosition-\bar{\knotPosition}_{k}\positionalQV\right)^{d}
\]
are recursively given by 
\begin{equation}
\bar{\approxCandCoeff}_{\left(-\lastKnotIndex-1\right)d}=0\label{eq:contribution-spline-start-formula}
\end{equation}
 and (\ref{eq:accumulation-rule}) for $k=-\lastKnotIndex,\dots,\lastKnotIndex$,
or, explicitly, 
\begin{align}
\bar{\approxCandCoeff}_{kd} & =\sum_{l=-\lastKnotIndex}^{k}\sum_{j=d}^{\degapprox}\binom{j}{d}\frac{}{}\bar{\hat{\approxCandCoeff}}_{lj}\left(\bar{\knotPosition}_{k}-\bar{\knotPosition}_{l}\right)^{j-d}\nonumber \\
 & =\bar{\hat{\approxCandCoeff}}_{kd}+\sum_{l=-\lastKnotIndex}^{k-1}\sum_{j=d}^{\degapprox}\binom{j}{d}\frac{}{}\bar{\hat{\approxCandCoeff}}_{lj}\left(\bar{\knotPosition}_{k}-\bar{\knotPosition}_{l}\right)^{j-d}\label{eq:contribution-piece-explicit}
\end{align}
 which follows from iteratively applying (\ref{eq:accumulation-rule}).
 Thereby, we aim for $\approxCandidate$ to come fairly close to $\approxFromLUT$.

\subsubsection{Conditions on Resulting Approximations}

So far, we have already assumed any contribution approximation $\approxCandidate$
to be continuous. If not, its definition (\ref{eq:contribution-spline-formula})
would be ambiguous at the knot positions $\bar{\knotPosition}_{k}\positionalQV$.

In addition to continuity, we also adopt the other conditions defining the set of eligible approximations $\approxKsctSet$ for normalized particle contributions, demanding
$\approxCandidate$ to have not more than $\numpapprox$ non-trivial
pieces, maximal polynomial degree $\degapprox$, compact support,
and be \textit{shifted even}, \ie fulfill 
\begin{equation}
\fc{\approxCandidate}{\knotPosition_{0}+\knotPosition}=\fc{\approxCandidate}{\knotPosition_{0}-\knotPosition}\label{eq:shifted-evenness}
\end{equation}
for some fixed $\knotPosition_{0}\in\mathbb{R}$ and all real $\knotPosition$.
Together with $\approxCandPiece{-\lastKnotIndex-1}\equiv\v 0$, shifted
evenness clearly implies compact support, \ie $\approxCandPiece{\lastKnotIndex}\equiv\v 0$.
This is particularly important because otherwise any non-zero coefficient
$\bar{\approxCandCoeff}_{\lastKnotIndex\;d}$ would add to regions
on the ray onto which the contribution is not meant to have any effect.
The error would be propagated at order $d$ even when using exact
integer arithmetic, annihilating all fruit of the quantization effort.

We don't expect requiring all these constraints to raise the approximation
error too much because the prototype $\approxFromLUT$ of $\approxCandidate$
already fulfills them, including shifted evenness with respect to
the ray parameter $\knotPosition_{\po}$. Still, as the quantization
operation disturbs all knot positions and coefficients, to an extend
dictated by the quantum values, we have to keep a close eye on this
additional source of error.

\subsubsection{Setting Quantized Knot Positions\label{subsec:Setting-Quantized-Knot-Positions}}

To minimize the overall shift between $\approxFromLUT$ and $\approxCandidate$,
we set the middle knot position quantum count to
\begin{equation}
\bar{\knotPosition}_{0}=\round{\frac{\knotPosition_{\po}}{\positionalQV}}\label{eq:quantized-middle-knot-position}
\end{equation}
where we use $\round{\cdot}$
to refer to just a usual nearest integer rounding function. In this
context it is irrelevant whether we round $z+\frac{1}{2}$ to $z$
or to $z+1$ for $z\in\mathbb{Z}$.

To keep the distances as close as possible to the non-quantized ones
and fulfill shifted evenness with respect to $\knotPosition_{0}=\bar{\knotPosition}_{0}\positionalQV$, we continue by setting 
\begin{equation}
\bar{\knotPosition}_{k}=\bar{\knotPosition}_{0}+\round{\frac{\pl\normalizedKnotPosition_{k}}{\positionalQV}}\text{ and }\;\bar{\knotPosition}_{-k}=\bar{\knotPosition}_{0}-\round{\frac{\pl\normalizedKnotPosition_{k}}{\positionalQV}}=2\bar{\knotPosition}_{0}-\bar{\knotPosition}_{k}\label{eq:quantized-knot-positions}
\end{equation}
for $k=1,\dots,\lastKnotIndex$, thus fulfilling $\bar{\knotPosition}_{k}-\bar{\knotPosition}_{0}=\bar{\knotPosition}_{0}-\bar{\knotPosition}_{-k}$
for all $k$.

\subsubsection{Evenness in Terms of Quantized Localized Difference Coefficients}

As we want the resulting particle contribution approximations to be shifted even as in (\ref{eq:shifted-evenness}), we want to express this condition in terms of the quantized difference coefficients.

Shifted evenness of $\approxCandidate$ is clearly equivalent to
\begin{equation}
\fc{\approxCandPiece k}{\knotPosition}=\fc{\approxCandPiece{-\left(k+1\right)}}{-\knotPosition}\text{ for }k=0,\dots,\lastKnotIndex\label{eq:inner-shifted-evenness}
\end{equation}
 which, together with (\ref{eq:quantized-localized-diff-coord-definition}),
implies 
\begin{align*}
\coefficientsQV\sum_{d=0}^{\degapprox}\bar{\hat{\approxCandCoeff}}_{kd}\left(\knotPosition-\bar{\knotPosition}_{k}\positionalQV\right)^{d} & =\fc{\approxCandPiece k}{\knotPosition}-\fc{\approxCandPiece{k-1}}{\knotPosition}\\
 & =-\left[\fc{\approxCandPiece{-k}}{-\knotPosition}-\fc{\approxCandPiece{-\left(k+1\right)}}{-\knotPosition}\right]\\
 & =-\coefficientsQV\sum_{d=0}^{\degapprox}\bar{\hat{\approxCandCoeff}}_{-kd}\left(-\knotPosition-\bar{\knotPosition}_{-k}\positionalQV\right)^{d}\\
 & =\coefficientsQV\sum_{d=0}^{\degapprox}\bar{\hat{\approxCandCoeff}}_{-kd}\left(-1\right)^{d+1}\left(\knotPosition-\bar{\knotPosition}_{k}\positionalQV\right)^{d},
\end{align*}
 hence,
\begin{equation}
\bar{\hat{\approxCandCoeff}}_{kd}=\left(-1\right)^{d+1}\bar{\hat{\approxCandCoeff}}_{-kd}\label{eq:contribution-spline-jumps-condition}
\end{equation}
for all $d=0,\dots,\degapprox$ and $k=0,\dots,\lastKnotIndex$.

In the special case of $k=0$, (\ref{eq:inner-shifted-evenness})
yields 
\begin{align*}
\coefficientsQV\sum_{d=0}^{\degapprox}\bar{\hat{\approxCandCoeff}}_{0d}\knotPosition^{d} & =\fc{\approxCandPiece 0}{\knotPosition}-\fc{\approxCandPiece{-1}}{\knotPosition}\\
 & =\fc{\approxCandPiece 0}{\knotPosition}-\fc{\approxCandPiece 0}{-\knotPosition}\\
 & =\coefficientsQV\sum_{d=0}^{\degapprox}\bar{\approxCandCoeff}_{0d}\left[\knotPosition^{d}-\left(-\knotPosition\right)^{d}\right],
\end{align*}
 hence, essentially,
\begin{equation}
\bar{\hat{\approxCandCoeff}}_{0d}=\begin{cases}
0 & \text{if }d\text{ is even}\\
2\bar{\approxCandCoeff}_{0d} & \text{else}.
\end{cases}\label{eq:contribution-spline-middle-jumps-condition}
\end{equation}

We have seen that (\ref{eq:inner-shifted-evenness}) implies the simple
conditions (\ref{eq:contribution-spline-jumps-condition}) and (\ref{eq:contribution-spline-middle-jumps-condition}).
And the converse is true, too, because (\ref{eq:contribution-spline-middle-jumps-condition})
directly provides (\ref{eq:inner-shifted-evenness}) for $k=0$ as
\begin{align*}
\fc{\approxCandPiece 0}{\knotPosition}-\fc{\approxCandPiece{-1}}{\knotPosition} & =\coefficientsQV\sum_{d=0}^{\degapprox}\bar{\hat{\approxCandCoeff}}_{0d}\knotPosition^{d}\\
 & =\coefficientsQV\sum_{d=0}^{\degapprox}\bar{\approxCandCoeff}_{0d}\left[\knotPosition^{d}-\left(-\knotPosition\right)^{d}\right]\\
 & =\fc{\approxCandPiece 0}{\knotPosition}-\fc{\approxCandPiece 0}{-\knotPosition},
\end{align*}
 and (\ref{eq:contribution-spline-jumps-condition}) yields 
\begin{align*}
\fc{\approxCandPiece k}{\knotPosition}-\fc{\approxCandPiece{k-1}}{\knotPosition} & =\coefficientsQV\sum_{d=0}^{\degapprox}\bar{\hat{\approxCandCoeff}}_{kd}\left(\knotPosition-\bar{\knotPosition}_{k}\positionalQV\right)^{d}\\
 & =\coefficientsQV\sum_{d=0}^{\degapprox}\bar{\hat{\approxCandCoeff}}_{-kd}\left(-1\right)^{d+1}\left(\knotPosition-\bar{\knotPosition}_{k}\positionalQV\right)^{d}\\
 & =-\coefficientsQV\sum_{d=0}^{\degapprox}\bar{\hat{\approxCandCoeff}}_{-kd}\left(-\knotPosition-\bar{\knotPosition}_{-k}\positionalQV\right)^{d}\\
 & =\fc{\approxCandPiece{-\left(k+1\right)}}{-\knotPosition}-\fc{\approxCandPiece{-k}}{-\knotPosition},
\end{align*}
 such that, by induction over $k$, 
\begin{align*}
\fc{\approxCandPiece k}{\knotPosition} & =\fc{\approxCandPiece{-\left(k+1\right)}}{-\knotPosition}-\fc{\approxCandPiece{-k}}{-\knotPosition}+\fc{\approxCandPiece{k-1}}{\knotPosition} \\
& =\fc{\approxCandPiece{-\left(k+1\right)}}{-\knotPosition}.
\end{align*}

\subsubsection{Setting Quantized Localized Difference Coefficients}

The above observations allows us to straightforwardly develop rules
for setting the quantized localized difference coefficients $\bar{\hat{\approxCandCoeff}}_{kd}$.

To obtain a function close to $\approxFromLUT$, we simply take all
coefficients contained in the look-up table entry, transform them
to correspond to $\approxFromLUT$, and round to nearest quantized
values. Specifically, we achieve
\begin{align*}
\frac{\coefficientsQV}{\positionalQV^{d}}\bar{\hat{\approxCandCoeff}}_{kd} & \approx\frac{\pma\sf\hat{\approxKsctCoeff}_{kd}}{\pd\pl^{d+3}}
\end{align*}
 by setting 
\begin{equation}
\bar{\hat{\approxCandCoeff}}_{kd}=\round{\frac{\positionalQV^{d}\pma\sf\hat{\approxKsctCoeff}_{kd}}{\coefficientsQV\pd\pl^{d+3}}}\label{eq:quantized-jumps-starting-from-lut}
\end{equation}
 for $k\ge1$, except for $k=1$ and uneven $d$ if $\numpapprox$
is uneven. The corresponding inversely located coefficients $\bar{\hat{\approxCandCoeff}}_{-kd}$
can then be simply computed as in (\ref{eq:contribution-spline-jumps-condition}).

The remaining coefficients are already determined by the shifted evenness condition and thus cannot be computed freely from look-up table entries. In case of
$\numpapprox$ being even, there is a middle knot with possibly non-zero
coefficients which have to obey (\ref{eq:contribution-spline-middle-jumps-condition}).
Thus, for uneven $d$,
\begin{align*}
\bar{\hat{\approxCandCoeff}}_{0d} & =2\bar{\approxCandCoeff}_{0d}\\
& =2\left[\bar{\hat{\approxCandCoeff}}_{0d}+\sum_{k=-\lastKnotIndex}^{-1}\sum_{j=d}^{\degapprox}\binom{j}{d}\frac{}{}\bar{\hat{\approxCandCoeff}}_{kj}\left(\bar{\knotPosition}_{0}-\bar{\knotPosition}_{k}\right)^{j-d}\right]
\end{align*}
 due to (\ref{eq:contribution-piece-explicit}), allowing us to compute
them as 
\begin{equation}
\bar{\hat{\approxCandCoeff}}_{0d}=-2\sum_{k=-\lastKnotIndex}^{-1}\sum_{j=d}^{\degapprox}\binom{j}{d}\frac{}{}\bar{\hat{\approxCandCoeff}}_{kj}\left(\bar{\knotPosition}_{0}-\bar{\knotPosition}_{k}\right)^{j-d}.\label{eq:quantized-middle-jumps-computation}
\end{equation}

Otherwise, there is no middle knot, \ie $\bar{\hat{\approxCandCoeff}}_{0d}=0$
for all $d$, but we still have to fulfill (\ref{eq:quantized-middle-jumps-computation}),
which is now equivalent to 
\begin{align*}
& \sum_{j=d}^{\degapprox}\binom{j}{d}\frac{}{}\bar{\hat{\approxCandCoeff}}_{-1j}\left(\bar{\knotPosition}_{0}-\bar{\knotPosition}_{-1}\right)^{j-d}\\
& =-\sum_{k=-\lastKnotIndex}^{-2}\sum_{j=d}^{\degapprox}\binom{j}{d}\frac{}{}\bar{\hat{\approxCandCoeff}}_{kj}\left(\bar{\knotPosition}_{0}-\bar{\knotPosition}_{k}\right)^{j-d}
\end{align*}
 for uneven $d$ and thus satisfied when setting
\begin{align}
\bar{\hat{\approxCandCoeff}}_{-1d} ={} & -\sum_{j=d+1}^{\degapprox}\binom{j}{d}\frac{}{}\bar{\hat{\approxCandCoeff}}_{-1j}\left(\bar{\knotPosition}_{0}-\bar{\knotPosition}_{-1}\right)^{j-d}\nonumber \\
& -\sum_{k=-\lastKnotIndex}^{-2}\sum_{j=d}^{\degapprox}\binom{j}{d}\frac{}{}\bar{\hat{\approxCandCoeff}}_{kj}\left(\bar{\knotPosition}_{0}-\bar{\knotPosition}_{k}\right)^{j-d}\nonumber \\
 ={} & -\sum_{k=-\lastKnotIndex}^{-2}\bar{\hat{\approxCandCoeff}}_{kd}\nonumber \\
 & -\sum_{j=d+1}^{\degapprox}\binom{j}{d}\frac{}{}\sum_{k=-\lastKnotIndex}^{-1}\bar{\hat{\approxCandCoeff}}_{kj}\left(\bar{\knotPosition}_{0}-\bar{\knotPosition}_{k}\right)^{j-d}\label{eq:quantized-jumps-uneven-pieces-count-computation}
\end{align}
in decreasing order of uneven $d$. Afterwards, the corresponding
values of $\bar{\hat{\approxCandCoeff}}_{1d}$ are again accessible
through (\ref{eq:contribution-spline-jumps-condition}).

As all involved values
during the computations in (\ref{eq:contribution-spline-jumps-condition}),
(\ref{eq:quantized-middle-jumps-computation}), and (\ref{eq:quantized-jumps-uneven-pieces-count-computation})
are integers, these operations can be carried out exactly, which
ensures the resulting approximation of the particle contribution to
be exactly shifted even.

\subsection{Derivation of Quantization Error Estimate $\QE$\label{appsec:quantization-error}}

We perform quantization for the very purpose of error elimination
but introduce with it a new source of error, hoping it will be less
severe.

The look-up table 
contains approximations $\approxKsct_{\normadst}$ to normalized particle contributions
$\ksct{\normadst}$ for several normalized distance values $\normadst$.
These approximations are optimized for given approximation parameters
$\numpapprox$ and $\degapprox$. These allow to generate optimal particle contribution
approximations $\approxFromLUT$, provided that the normalized distance
of the particle to the viewing ray is equal to $\adst$
for a table entry. While this is not the case in general, the error resulting from a deviation in $\adst$ is easily mitigated by providing a large
number of entries in the look-up table. We therefore neglect this error and
assume $\approxFromLUT$ to be almost optimal.

Yet, we don't utilize $\approxFromLUT$ directly but convert it into
a quantized version $\approxCandidate$ following the steps explained
in \Section \ref{appsec:quantized-knots} thereafter.
This involves rounding knot positions to full multiples of the quantum
value $\positionalQV$ and setting all localized difference coefficients
to multiples of a second quantum value $\coefficientsQV$. This
distorts the “almost optimal” approximation $\approxFromLUT$
in a way that we cannot ignore because the error introduced here—which
we call the \emph{quantization error}—depends primarily on the quantum
values $\positionalQV$ and $\coefficientsQV$. Keeping the quantization
error small by setting these to arbitrarily small values is not an
option as this would increase the integers $\bar{\knotPosition}_{k}$,
$\bar{\hat{\approxCandCoeff}}_{kd}$, and $\bar{\approxCandCoeff}_{kd}$
by the same factor, making them costly to operate on.

To be able to balance all involved parameters, we seek to compute
the quantization error. While we can directly compute it for a specific
instance of a quantized contribution approximation, we cannot simply generalize this
because just a small shift on a viewing ray or in particle parameters
can change the quantization error dramatically. The error may even
vanish if quantum multiples are directly hit by chance. Therefore,
we have to tolerate our generalized quantization error measure to
be a more vague estimation. Thereby, we attempt to introduce as little
bias as possible, estimating the expected error as opposed to providing
an upper bound.

As with the approximation error, we start off by assuming the contributing
particle to be normalized, \ie to carry unit mass, density, smoothing
radius, and field attribute. Our quantization error estimation has
two components, one measuring the error by quantization of knot positions,
which depends on the length quantum $\positionalQV$, the other one
by quantization of the difference coefficients, based on the value
quantum $\coefficientsQV$.

\subsubsection{Positional Quantization Error}

Since the knots defining $\approxFromLUT$ can be located at just
any ray parameter, we slightly shift them as described in \Section
\ref{subsec:Setting-Quantized-Knot-Positions} to guarantee that they
are multiples of the length quantum $\positionalQV$. The effect is
similar to a slight shift of the approximation on the ray. The shifting
amount depends on how close the original positions are to multiples
of $\positionalQV$. Based on the fact that the distance to the nearest
multiple of $\positionalQV$ is an element of the interval $\left[0,\frac{\positionalQV}{2}\right]$
and assuming a uniform probability distribution for the selection
from this interval, we simply let the shift be $\frac{\positionalQV}{4}$
to obtain a suitable mean error. More precisely, we expect the average
error caused by knot position quantization to be similar to the error
of shifting the contribution model $\approxFromLUT$ taken from the
look-up table by $\frac{\positionalQV}{4}$ along the ray. Measuring
this error by its $L^{2}$ norm, it amounts to 
\[
\left(\intop_{\mathbb{R}}\left[\fc{\approxFromLUT}{\knotPosition}-\fc{\approxFromLUT}{\knotPosition-\frac{\positionalQV}{4}}\right]^{2}\diff\knotPosition\right)^{\frac{1}{2}}.
\]
 However, we don't compute this error directly but apply further approximations.
Firstly, since we hope $\approxFromLUT$ to resemble the actual normalized particle contribution $\ksct{\normadst}$ fairly well, we can expect the above error
measure to stay approximately equal when replacing $\approxFromLUT$
by $\ksct{\normadst}$. Thereby, our error estimation becomes independent
from the actually used piecewise polynomial approximation.

Secondly, to make the error estimate easier computable, we approximate
the shifted kernel section by a first-order expression, specifically
\[
\fc{\ksct{\adst}}{\knotPosition-\frac{\positionalQV}{4}}\approx\fc{\ksct{\adst}}{\knotPosition}-\frac{\positionalQV}{4}\cdot\fc{\frac{\diff}{\diff\knotPosition}\ksct{\adst}}{\knotPosition},
\]
 yielding the error estimate of 
\[
\frac{\positionalQV}{4}\left(\intop_{\mathbb{R}}\left[\fc{\frac{\diff}{\diff\knotPosition}\ksct{\adst}}{\knotPosition}\right]^{2}\diff\knotPosition\right)^{\frac{1}{2}}.
\]
To become independent from $\adst$ and the screen resolution, we integrate this per-ray measure over all viewing rays in one direction,
which results in $\frac{\positionalQV}{4}\knormSpecial$,
where we abbreviate
\[
\knormSpecial=\left(\intop_{0}^{\kSubBorder}\adst\intop_{\mathbb{R}}\left[\fc{\frac{\diff}{\diff\knotPosition}\ksct{\adst}}{\knotPosition}\right]^{2}\diff\knotPosition\diff\adst\right)^{\frac{1}{2}},
\]
which for our example cubic B-spline kernel (\ref{eq:cubic-kernel}) happens to amount
to 
\[
\knormSpecial=\frac{\sqrt{14}}{7\pi}\approx0.170\text{ .}
\]
Ultimately, to obtain a relative error measure, we relate it to the $L^2$  norm of the normalized particle contribution itself. Abbreviating it by
\[
\knorm = \left(4\pi\int\limits _{t=0}^{\infty}\left[t\fc{\ks}{t}\right]^{2}\diff t\right)^{\frac{1}{2}},
\]
which is roughly $0.352$ for the cubic B-spline kernel,
we arrive at the positional quantization error component estimate
\[
\fc{\positionalQE}{\positionalQV}=\frac{\positionalQV\knormSpecial}{4\knorm}
\]

\subsubsection{Coefficients Quantization Error}

The second quantization error component is aimed at estimating the
effect of quantizing the localized difference coefficients $\hat{\approxKsctCoeff}_{kd}$.
At every knot position, these difference coefficients constitute the
values by which the corresponding localized coefficients $\approxKsctCoeff_{kd}$
of the normalized contribution approximation “jump”. Therefore, a small shift of the difference
coefficients $\hat{\approxKsctCoeff}_{kd}$ as if following (\ref{eq:quantized-jumps-starting-from-lut})
for a normalized particle causes the exact same shift on the localized
coefficients $\approxKsctCoeff_{kd}$ at the knot position for the
same order $d$. Given the quantum value $\coefficientsQV_d=\frac{\coefficientsQV}{\positionalQV^d}$, the
shift when rounding to a nearest full quantum multiple is within the
interval $\left[0,\coefficientsQV_d\right]$, selected at a uniform
probability distribution. We follow the same strategy as for the positional
quantization case, assuming the shift to amount to $\frac{1}{4}\coefficientsQV_d$
to obtain a suitable mean error value. When applied to the localized
difference coefficient of order $d$ at knot positions $\normalizedKnotPosition_{k}$,
this causes the approximation to differ by 
\begin{align}
& \left(\approxKsctCoeff_{kd}+\frac{\coefficientsQV_d}{4}\right)\left(\knotPosition-\normalizedKnotPosition_{k}\right)^{d}-\approxKsctCoeff_{kd}\left(\knotPosition-\normalizedKnotPosition_{k}\right)^{d}\nonumber \\
& =\frac{\coefficientsQV_d}{4}\left(\knotPosition-\normalizedKnotPosition_{k}\right)^{d}\label{eq:coeff-quantization-difference}
\end{align}
 for $\knotPosition>\normalizedKnotPosition_{k}$. Consequently, the
error of coefficient quantization is independent from the coefficient
themselves, only depending on the order $d$. It is thus independent
from the model to be quantized and, in fact, from the SPH kernel function
used.

Further, (\ref{eq:coeff-quantization-difference}) shows an error
of order $d$, analogous to the higher-order errors caused by rounding,
which we are trying to eliminate in the first place. For $d>0$, these
distortions grow along the ray, until the next knot at position $\normalizedKnotPosition_{k+1}$
is reached and the new difference coefficient is added to the coefficient
localized there. This difference coefficient is also distorted by
quantization but this distortion has the same probability of adding
to the original error as of subtracting from it. To obtain a suitable
mean estimate of the error to be expected, we assume the error to
stay the same, \ie the difference to the original model to be 
\[
\frac{\coefficientsQV_d}{4}\left(\knotPosition-\normalizedKnotPosition_{k}\right)^{d}
\]
 also for $\knotPosition>\normalizedKnotPosition_{k+1}$. Consequently,
for our error estimate we ignore intermediate knots and only consider
the impact of coefficient quantization in the first knot modeling
a particle contribution.

In contrast to the situation for the higher-order errors through rounding
localized floating-point coefficients, the distance upon which the
higher-order quantization errors can operate on is restricted. As
we set the innermost difference coordinates guaranteeing the quantized
particle contribution approximation to be shifted-even (see \Section \ref{appsec:quantized-knots}), they only
act on the distance from an outermost knot position to the center
of the contribution approximation on the viewing ray. In other words,
the error has a bounded support of length $2\normalizedKnotPosition_{\lastKnotIndex}$
and is thus $L^{2}$ measurable, its norm being 
\begin{align*}
& \left(\intop_{-\normalizedKnotPosition_{\lastKnotIndex}}^{0}\left[\frac{\coefficientsQV_d}{4}\left(\knotPosition-\normalizedKnotPosition_{-\lastKnotIndex}\right)^{d}\right]^{2}\diff\knotPosition+\intop_{0}^{\normalizedKnotPosition_{\lastKnotIndex}}\left[\frac{\coefficientsQV_d}{4}\knotPosition^{d}\right]^{2}\diff\knotPosition\right)^{\!\frac{1}{2}} \\
& =\frac{\coefficientsQV_d}{4}\left(2\intop_{0}^{\normalizedKnotPosition_{\lastKnotIndex}}\knotPosition^{2d}\diff\knotPosition\right)^{\frac{1}{2}}\\
 & =\frac{\coefficientsQV_d}{4}\sqrt{\frac{2\normalizedKnotPosition_{\lastKnotIndex}^{2d+1}}{2d+1}}.
\end{align*}
Assuming the outermost knot positions to approximate the boundaries of the
particle volume of influence, \ie 
\[
\normalizedKnotPosition_{\lastKnotIndex}\approx\sqrt{\kSubBorder^{2}-\adst^{2}},
\]
 allows us to become independent from any aspects of the contribution approximation $\bestApproxKsct$ except for maximal degree $\degapprox$, expecting
an error of 
\[
\frac{\coefficientsQV_d}{4}\sqrt{\frac{2\sqrt{\kSubBorder^{2}-\adst^{2}}^{2d+1}}{2d+1}}
\]
for quantizing the difference coefficients of order $d\le\degapprox$. Here, $\kSubBorder$ denotes the upper bound of the SPH kernel function's support.

Integrating over all viewing rays and relating the result to the $L^2$ norm $\knorm$ of the normalized particle contribution in analogy to the positional quantization error component estimate $\positionalQE$,
we define the coefficient quantization error estimate of order $d$
to be
\begin{align*}
\fc{\coefficientsQE d}{\coefficientsQV_d} & =\frac{\coefficientsQV_d}{4\knorm}\left(\frac{2}{2d+1}\intop_{0}^{\kSubBorder}\adst\cdot\sqrt{\kSubBorder^{2}-\adst^{2}}^{2d+1}\diff\adst\right)^{\frac{1}{2}}\\
 & =\frac{\coefficientsQV_d}{4\knorm}\left(\left[-\frac{2\sqrt{\kSubBorder^{2}-\adst^{2}}^{2d+3}}{\left(2d+1\right)\left(2d+3\right)}\right]_{\adst=0}^{\kSubBorder}\right)^{\frac{1}{2}}\\
 & =\frac{\coefficientsQV_d}{4\knorm}\sqrt{\frac{2\kSubBorder^{2d+3}}{\left(2d+1\right)\left(2d+3\right)}},\\
\intertext{\text{amounting to}}\fc{\coefficientsQE d}{\coefficientsQV_d} & =\frac{2^{d}\coefficientsQV_d}{\knorm\sqrt{\left(2d+1\right)\left(2d+3\right)}}
\end{align*}
 for $\kSubBorder=2,$ such as in the case of the cubic kernel
(\ref{eq:cubic-kernel}).

\subsubsection{Combined Quantization Error Estimate}

Now we want to generate a combined quantization error estimate, integrating
the components $\positionalQE$ and $\coefficientsQE 0,\dots,\coefficientsQE{\degapprox}$.
We could simply sum up all of these but this would introduce an overestimation
bias as it would model all field distortions acting in the same direction.
Each error component is the $L^{2}$ norm of an $L^{2}$-measurable
field $\mathbb{R}^{3}\to\mathbb{R}$ modeling the alteration of the
SPH attribute field by the respective quantization step. An ideal
combination of the error components would be the $L^{2}$ norm of
the sum of these alterations. We therefore combine the components
as if the underlying difference fields were orthogonal with respect
to the inner product inducing the $L^{2}$ norm, \ie as if 
\[
\left\langle \fieldA,\fieldB\right\rangle =\intop_{\mathbb{R}^{3}}\fc{\fieldA}{\v x}\fc{\fieldB}{\v x}\diff\v x=0
\]
for any two such difference fields $\fieldA$ and $\fieldB$ measured
by different error components. While we do not assume this to be the
case, we still expect the resulting error measure to be much more
realistic. Instead of summing the components, this model demands to
sum their squares and take the square root afterwards. Employing $\coefficientsQV_d=\frac{\coefficientsQV}{\positionalQV^d}$, we therefore propose
the quantization error estimate
\begin{align*}
\fc{\QE}{\positionalQV,\coefficientsQV} & =\sqrt{\fc{\positionalQE^{2}}{\positionalQV}+\sum_{d=0}^{\degapprox}\fc{\coefficientsQE d^{2}}{\frac{\coefficientsQV}{\positionalQV^{d}}}}\\
 & =\frac{1}{4\knorm}\sqrt{\knormSpecial^{2}\positionalQV^{2}+\sum_{d=0}^{\degapprox}\frac{2\kSubBorder^{2d+3}}{\left(2d+1\right)\left(2d+3\right)}\cdot\frac{\coefficientsQV^{2}}{\positionalQV^{2d}}}.
\end{align*}

While $\fc{\QE}{\positionalQV,\coefficientsQV}$ can only be a rough
estimate of the error to be expected, its basic properties are plausible.
For fixed length quantum $\positionalQV$, it grows with $\coefficientsQV$
asymptotically linearly because the positional quantization component
$\positionalQE$ is invariant in $\coefficientsQV$ and all other
components are linear in $\coefficientsQV$. The exact quantization
error can be expected to show a very similar behavior. After all,
smaller value quantum values can only reduce the discrepancy of the
quantized model from its original version, albeit at the cost of increasing
the involved integers representing the multiples.

The behavior with respect to $\positionalQV$ is reasonable, as well.
While the positional quantization component grows linearly with $\positionalQV$
as expected, the value quantization components of order $d>0$ decrease
and even approach zero. This is easily explained by the fact that
a larger length quantum $\positionalQV$ directly means a smaller
value quantum $\coefficientsQV_d=\frac{\coefficientsQV}{\positionalQV^d}$ for positive order $d$. This correlation also
accounts for the value quantization errors to approach infinity for
$\positionalQV\to0$.

\subsection{Minimization of \(\QE\)\label{appsec:quantization-error-minimization}}

In case of $\degapprox>0$ and for fixed $\coefficientsQV$ and
$,\dots,$,
the combined quantization error estimate $\fc{\QE}{\positionalQV,\coefficientsQV}$,
$\positionalQV\in\left(0,\infty\right)$ for a normalized particle
satisfies 
\begin{align*}
\lim_{\positionalQV\to0}\fc{\QE}{\positionalQV,\coefficientsQV} & \ge\lim_{\positionalQV\to0}\fc{\coefficientsQE 1}{\frac{\coefficientsQV}{\positionalQV}}=\infty,\\
\lim_{\positionalQV\to\infty}\fc{\QE}{\positionalQV,\coefficientsQV} & \ge\lim_{\positionalQV\to\infty}\fc{\positionalQE}{\positionalQV}=\infty.
\end{align*}
 This proves that the first partial derivative of the square of $\fc{\QE}{\positionalQV,\coefficientsQV}$
with respect to $\positionalQV$, 
\begin{align*}
& \fc{\frac{\diff}{\diff\positionalQV}\QE^{2}}{\positionalQV,\coefficientsQV}\\
&=\frac{\knormSpecial^{2}\positionalQV}{8\knorm^2}-\sum_{d=1}^{\degapprox}\frac{d\kSubBorder^{2d+3}}{4\knorm^2\left(2d+1\right)\left(2d+3\right)}\cdot\frac{\coefficientsQV^{2}}{\positionalQV^{2d+1}},
\end{align*}
 attains both negative and positive values within $\positionalQV\in\left(0,\infty\right)$.
Moreover, $\fc{\frac{\diff}{\diff\positionalQV}\QE^{2}}{\positionalQV,\coefficientsQV}$
is strictly growing with $\positionalQV$ because
\[
\fc{\frac{\diff^{2}}{\diff\positionalQV^{2}}\QE^{2}}{\positionalQV,\coefficientsQV}=\frac{\knormSpecial^{2}}{8\knorm^2}+\sum_{d=1}^{\degapprox}\frac{d\kSubBorder^{2d+3}}{4\knorm^2\left(2d+3\right)}\cdot\frac{\coefficientsQV^{2}}{\positionalQV^{2d+2}}>0
\]
 for all positive $\coefficientsQV$ and $\positionalQV$.

In consequence, $\fc{\QE}{\positionalQV,\coefficientsQV}$ attains
one global minimum at \emph{the} optimal length quantum, which can be computed numerically by running a root-finding
method on $\fc{\frac{\diff}{\diff\positionalQV}\QE^{2}}{\positionalQV,\coefficientsQV}$.

\end{multicols}
\end{document}